\documentclass[12pt,preprint]{aastex}

\def\R#1{{\mathrm{#1}}}
\def\ltsima{$\; \buildrel < \over \sim \;$}
\def\simlt{\lower.5ex\hbox{\ltsima}}
\def\gtsima{$\; \buildrel > \over \sim \;$}
\def\simgt{\lower.5ex\hbox{\gtsima}}
\shorttitle{RAVE 1st data release}
\shortauthors{Steinmetz et al.}

\begin{document}


\title{The Radial Velocity Experiment (RAVE): first data release}
\author{
M. Steinmetz\altaffilmark{1},
T. Zwitter\altaffilmark{2},
A. Siebert\altaffilmark{1,3},
F.~G. Watson\altaffilmark{4},
K.~C. Freeman\altaffilmark{5},
U. Munari\altaffilmark{6},
R. Campbell\altaffilmark{7},
M. Williams\altaffilmark{5},
G.~M. Seabroke\altaffilmark{8},
R.~F.~G. Wyse\altaffilmark{9},
Q.~A. Parker\altaffilmark{7,4},
O. Bienaym\'e\altaffilmark{10},
S. Roeser\altaffilmark{11},
B.~K. Gibson\altaffilmark{12},
G. Gilmore\altaffilmark{8},
E.~K. Grebel\altaffilmark{13},
A. Helmi\altaffilmark{14},
J.~F. Navarro\altaffilmark{15},
D. Burton\altaffilmark{4},
C.~J.~P. Cass\altaffilmark{4},
J.~A. Dawe\altaffilmark{\dagger,4},
K. Fiegert\altaffilmark{4},
M. Hartley\altaffilmark{4},
K.~S. Russell\altaffilmark{4},
W. Saunders\altaffilmark{4},
H. Enke\altaffilmark{1},
J. Bailin\altaffilmark{16},
J. Binney\altaffilmark{17},
J. Bland-Hawthorn\altaffilmark{4},
C. Boeche\altaffilmark{1},
W. Dehnen\altaffilmark{18},
D.~J. Eisenstein\altaffilmark{3},
N.~W. Evans\altaffilmark{8},
M. Fiorucci\altaffilmark{6},
J.~P. Fulbright\altaffilmark{11},
O. Gerhard\altaffilmark{13},
U. Jauregi\altaffilmark{2},
A. Kelz\altaffilmark{1},
L. Mijovi\'{c}\altaffilmark{2},
I. Minchev\altaffilmark{19},
G. Parmentier\altaffilmark{8},
J. Pe\~narrubia\altaffilmark{15},
A.~C. Quillen\altaffilmark{19},
M.~A. Read\altaffilmark{20},
G. Ruchti\altaffilmark{11},
R.-D. Scholz\altaffilmark{1},
A. Siviero\altaffilmark{6},
M.~C. Smith\altaffilmark{14},
R. Sordo\altaffilmark{6},
L. Veltz\altaffilmark{12},
S. Vidrih \altaffilmark{8},
R. von Berlepsch\altaffilmark{1},
B.~J. Boyle\altaffilmark{21},
E. Schilbach\altaffilmark{11}
}

\altaffiltext{1}{Astrophysikalisches Institut Potsdam, Potsdam, Germany}
\altaffiltext{2}{University of Ljubljana, Department of Physics, Ljubljana, Slovenia}
\altaffiltext{3}{Steward Observatory, Tucson AZ, USA}
\altaffiltext{4}{Anglo Australian Observatory, Sydney, Australia}
\altaffiltext{5}{RSAA, Mount Stromlo Observatory, Canberra, Australia}
\altaffiltext{6}{INAF Osservatorio Astronomico di Padova, Asiago, Italy}
\altaffiltext{7}{Macquarie University, Sydney, Australia}
\altaffiltext{8}{Institute of Astronomy, University of Cambridge, UK}
\altaffiltext{9}{Johns Hopkins University, Baltimore MD, USA}
\altaffiltext{10}{Observatoire de Strasbourg, Strasbourg, France}
\altaffiltext{11}{Astronomische Rechen Institut, Heidelberg, Germany}
\altaffiltext{12}{University of Central Lancashire, Preston, UK}
\altaffiltext{13}{Astronomical Institute of the University of Basel, Basel, Switzerland}
\altaffiltext{14}{Kapteyn Astronomical Institute, University of Groningen, Groningen, The Netherlands}
\altaffiltext{15}{University of Victoria, Victoria, Canada}
\altaffiltext{16}{Centre for Astrophysics and Supercomputing, Swinburne University
	of Technology, Hawthorn, Australia}
\altaffiltext{17}{Rudolf Peierls Centre for Theoretical Physics, University of Oxford, UK}
\altaffiltext{18}{University of Leicester, Leicester, UK}
\altaffiltext{19}{University of Rochester, Rochester NY, USA}
\altaffiltext{20}{University of Edinburgh, Edinburgh, UK}
\altaffiltext{21}{Australia Telescope National Facility, Sydney, Australia}
\altaffiltext{$\dagger$}{Deceased. This paper is dedicated to the memory of John Alan Dawe (1942-2004), Astronomer-in-Charge of the UK Schmidt Telescope 1978-1984, and enthusiastic RAVE observer 2003-4.}



\begin{abstract}
We present the first data  release of the Radial Velocity Experiment (RAVE),
an ambitious  spectroscopic survey to measure radial  velocities and stellar
atmosphere parameters  (temperature, metallicity, surface gravity)  of up to
one million  stars using the 6dF  multi-object spectrograph on  the 1.2-m UK
Schmidt  Telescope  of the  Anglo-Australian  Observatory  (AAO).  The  RAVE
program  started  in  2003,  obtaining  medium  resolution  spectra  (median
R=7,500) in  the Ca-triplet  region ($\lambda\lambda$ 8,410--8,795  \AA) for
southern hemisphere  stars drawn from the Tycho-2  and SuperCOSMOS catalogs,
in the magnitude range  $9\!\!<\!\!I\!\!<\!\!12$.  The first data release is
described in this paper and contains radial velocities for 24,748 individual
stars (25,274 measurements when including re-observations).  Those data were
obtained on 67 nights between 11 April 2003 to 03 April 2004.  The total sky
coverage  within  this data  release  is  $\sim$4,760  square degrees.   The
average signal to  noise ratio of the observed spectra is  29.5, and 80\% of
the radial  velocities have  uncertainties better than  3.4~km/s.  Combining
internal  errors and  zero-point errors,  the mode  is found  to  be 2~km/s.
Repeat observations are used to  assess the stability of our radial velocity
solution,  resulting in  a variance  of 2.8~km/s.   We demonstrate  that the
radial velocities derived for the first  data set do not show any systematic
trend  with  color or  signal  to noise.   The  RAVE  radial velocities  are
complemented  in the  data release  with  proper motions  from Starnet  2.0,
Tycho-2  and SuperCOSMOS,  in addition  to photometric  data from  the major
optical and infrared catalogs (Tycho-2,  USNO-B, DENIS and 2MASS).  The data
release can be accessed via the RAVE webpage: http://www.rave-survey.org.
\end{abstract}



\section{Introduction}
\label{s:introduction}

Within the past decade it is  being increasingly recognised that many of the
clues to the  fundamental problem of galaxy formation  in the early Universe
are contained in the motions and chemical composition of long-lived stars in
our Milky Way galaxy (see  e.g. Freeman \& Bland-Hawthorn 2002).  The recent
discovery of several instances of  tidal debris in our Galaxy challenges the
view laid down in the seminal  paper by \citet{els}, who envision the Galaxy
to be formed in one major monolithic collapse at an early epoch, followed by
a period of  relative quiescence, lasting many Gyr.   These examples include
the discovery of the tidally distorted/disrupted Sagittarius dwarf galaxy
\citep{sgr}, the photometrically identified low-latitude Monoceros structure
in  the Sloan  Digital Sky  Survey  \citet{monoceros} and  the multitude  of
features in higher latitude fields \citep{belokurov}.

\clearpage

\begin{figure*}[hbtp]
\centering
\includegraphics[width=8.7cm,angle=270]{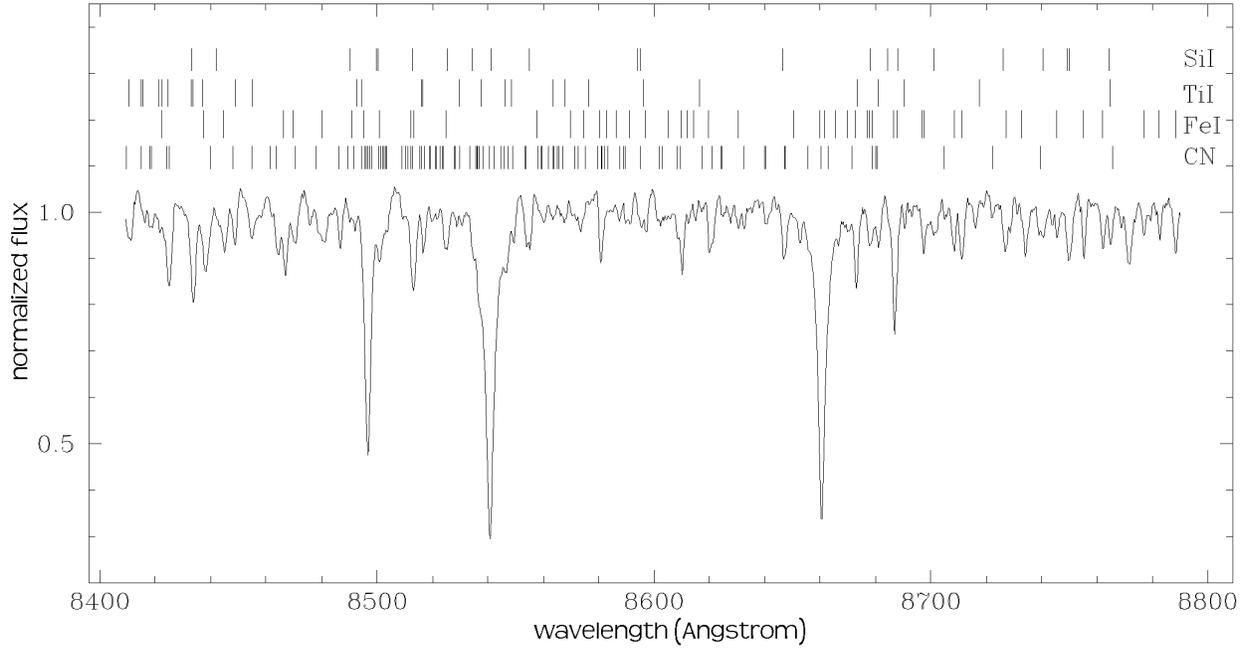}
\caption{The 
RAVE spectrum of a typical  field star, HD~154837 (K0~III), illustrating the
properties  of   the  chosen  wavelength   interval  around  the   Ca~II  IR
triplet. The strongest other absorption lines are identified.  }
\label{f:nicespectrum}
\end{figure*}

\clearpage

Furthermore,  within   the  context   of  the  concordance   LCDM  scenario,
sophisticated computer simulations of structure growth within a CDM universe
have now  begun to shed light on  how the galaxy formation  process may have
taken  place in  a hierarchical  framework (see  e.g.  Steinmetz  \& Navarro
2002,  Abadi  et  al.~2003,   Brook  et  al.~2005,  Governato  et  al.~2004,
Sommer-Larsen et  al.~2003).  These analyses  lead to a  reinterpretation of
structures such  as the Eggen  moving groups \citep{navarro04} or  the Omega
Cen globular cluster  \citep{meza05} in terms of merger  remnants.  In fact,
in the  extreme case, structures, and  old stars, in even  the thin Galactic
disk  are   attributed  to  accretion   events  \citep{abadi03,meza05}.   In
addition,  strong  evidence  for  accretion and  assimilation  of  satellite
galaxies  can  also be  seen  for  other galaxies  in  the  Local Group,  in
particular the  `great stream'  in M31 \citep{sgrstream}.   However, whether
the Galaxy can indeed be formed by a sequence of merging events as predicted
by  current cosmological  models of  galaxy  formation, or  whether the  few
well-established accretion and merging remnants  -- which account for only a
small fraction of the stellar mass of the Galaxy -- are all there is, is still a
largely  unanswered question.  Large  kinematic surveys  are needed,  as are
large  surveys  that  derive  chemical  abundances,  since  both  kinematics
signatures and elemental abundance signatures persist longer than do spatial
over-densities.  It  is still  unclear (Wyse \&  Gilmore 2006),  whether the
observed chemical properties of the stars in the thick disk (Gilmore, Wyse
\& Jones 1995)  can be brought into agreement with a  scenario that sees the
thick disk primarily  as the result of accretion  events \citep{abadi03}.  A
similar  question arises owing  to the  distinct age  distribution (Unavane,
Wyse \& Gilmore 1996, see however  Abadi et al.~2003) and chemical elemental
abundance distributions of  stars in the stellar halo and  in low-mass dwarf
galaxies of the Local Group  (Tolstoy et al.~2003; see however, Robertson et
al.~2005, Bullock \& Johnston 2005).

Stellar clusters, spiral arms, and the  Galactic bar leave an imprint in the
chemical   and  stellar   velocity  distribution  in   the  solar
neighborhood (Dehnen 2000; Quillen \&  Minchev 2005; de Simone et al.~2004)
as well. Multidimensional databases are required to  investigate  and
differentiate  between these  processes  and structure  caused by  satellite
accretion.

The growing awareness of the importance  of the `fossil record' in the Milky
Way  Galaxy in  constraining galaxy  formation  theory is  reflected by  the
increasing number of  missions designed to unravel the  formation history of
the Galaxy.  Stellar spectroscopy plays a crucial role in these studies, not
only providing  radial velocities  as a key  component of  the 6-dimensional
phase  space  of  stellar  positions  and  velocities,  but  also  providing
much-needed  information   on  the  gravity  and   chemical  composition  of
individual stars.  An example of  the power of such multidimensional stellar
datasets  have  recently  been  shown  by  \citet{helmi06}  who,  by  using  a
combination of proper motions and distances from the Hipparcos catalog
\citep{hip} and  spectra from  the Geneva-Copenhagen Survey  (Nordstr\"om et
al.~ 2005,  hereafter  GCS)  were  able to  identify  several  accretion
candidates within the immediate neighborhood of the Sun.

However, despite  the importance of  stellar spectroscopy, the  past decades
have seen  only limited progress. Soon after  \citet{vogel1873} measured the
radial  velocities  of  Sirus  and  Procyon,  \citet{oldgeorge}\footnote{The
great-great-grandfather of G. Seabroke,  co-author of this paper.} performed
one of the first surveys,  measuring 68 radial velocities for  29 stars, followed by
699 observations of  40 stars \citep{oldgeorge2} and 866  observations of 49
stars  \citep{oldgeorge3}.  Since  then,  over the  next  125 years,  radial
velocities  for  some 50,000  stars  have  become  available in  the  public
databases of the Centre  de Donn\'ees astronomiques de Strasbourg (hereafter
CDS).   This is  surprisingly few,  compared to  the more  than  one million
galaxy redshifts  measured within the  past decade.  This sample  of stellar
radial  velocities has  recently been  increased substantially  by  the GCS,
containing radial velocities for 16,682 nearby dwarf stars, and by
\citet{famay} publishing 6,691 radial velocities for apparently bright giant
stars.  Both catalogs were part of the Hipparcos follow-up campaign.

With the advent of wide  field multi-object spectroscopy (MOS) fiber systems
in the  1990's, pioneered  particularly at the  AAO with FOCAP,  AUTOFIB and
most  recently  with  the 2dF  and  6dF  instruments  on  the AAT  and  UKST
respectively  (e.g.   Lewis  et  al.~  2002 and  Watson  et  al.~2000),  the
possibility  of  undertaking  wide-area  surveys with  hemispheric  coverage
became  feasible.   Initially,  such   projects  were  more  concerned  with
large-scale galaxy  and quasar redshift  surveys (e.g.  Colless  et al.~2001
for 2dF and  Jones et al.~2004 for 6dF).  Apart from  the samples of several
hundred  to a  few thousand  stars obtained  prior to  the  commissioning of
AAOmega at the AAT (see for example Kuijken \& Gilmore 1989a, b, c; Gilmore,
Wyse \& Jones  1995; Wyse \& Gilmore 1995; Gilmore, Wyse  \& Norris 2002) no
large-scale, wide area stellar  spectroscopy projects had been undertaken in
our own  galaxy.  This  has now changed  with new surveys  like SDSSII/SEGUE
already under way, with a planned delivery of 240,000 spectra by mid 2008
\citep{segue}, and the capabilities of  AAOmega on the AAT.  Over a slightly
longer time frame,  the RAdial Velocity Experiment (RAVE), the survey we describe in this paper,  is expected to provide spectra for  up to 1 Million
stars by 2011.  This trend for large stellar surveys will culminate with the
ESA cornerstone mission Gaia, which, in addition to astrometric information,
will provide  multi-epoch radial  velocities for up  to one  hundred million
stars, by  2018. Each of these surveys  has its own unique  aspect, and they
are largely complementary in capabilities and target sample.

With  a radial  velocity error  of about  2~km/s and  80\%  of measurements
better than 3.4~km/s, the RAVE velocities are accurate enough for almost any
Galactic kinematical study.   Radial velocities are however just  one of the
necessary  stellar  parameters:   proper  motions,  distances  and  chemical
abundances  are  also  needed.   Proper  motions  of  varying  accuracy  are
available  for most  of  the RAVE  stars,  via Starnet2.0,  Tycho-2 or  SSS.
Already, some  of the  cooler dwarfs ($J-K>0.5$)  with more  accurate proper
motions can be identified as  dwarfs from their reduced proper motions.  For
these  stars, it is  possible to  estimate all  six phase  space coordinates
using their photometric parallaxes.

For most of these stars,  there is no previous spectroscopic information, so
the RAVE sample provides many scientific opportunities.  Some of the science
programs in progress by RAVE team members include:
\begin{itemize} 
\item Discovery of extreme velocity stars and estimates of the local escape
  velocity and total mass of the Galaxy
\item The 3D velocity distribution function of the local Galactic disk
\item Kinematics of the main stellar components of the Galaxy
\item Characterization of the local Galactic disk  potential and the structure of the disk components
\item Substructure in the disk and halo of the Galaxy, including the 
  Arcturus, Sagittarius and other star streams
\item Elemental abundances of high velocity stars
\item Calibration of stellar atmospheric parameters and correspondence 
  with the MK scheme through the HR diagram 
\item Searches for spectroscopic binaries and cataclysmic variables
\item The $\lambda8,620$\AA~diffuse interstellar band as an estimator of 
  interstellar reddening.
\end{itemize}

In this  paper we describe the first  data release of the  RAVE survey which
contains radial velocities obtained  from RAVE spectra (the spectra, stellar
parameters and additional  information will be part of  the further releases
as  the  first  year  spectra  are  contaminated  by  second  order  light).
Photometric and proper motion data  from other surveys are also provided for
ease of  use. The  structure of  the paper is  as follows:  In Section  2 we
describe the survey layout,  technical equipment and input catalog.  Section
3 is devoted to the actual observations, followed by a section detailing the
data reduction. Section 5 discusses  the data quality and compares RAVE data
with  independent data  taken  with other  telescopes.   Finally, Section  6
provides  a detailed  description  of the  data  product of  the first  data
release (henceforth DR1) and concludes with longer term perspectives.

\section{Survey design and input catalog}
\label{s:survey_design_main}

The wide field-of-view of the  UK Schmidt and the multiplexing capability of
6dF are well-matched to a survey of apparently bright stars.  The scientific
goals of RAVE include analysing  the chemical and dynamical evolution of the
Galaxy,  using as  tracers both  dwarfs and  giants observed  locally.  Most
apparently bright  stars will  be in  the thin and  thick disks;  adopting a
limiting magnitude of  $I\! \sim \! 12$ (see  below), dwarfs probe distances
of hundreds  of parsec  and giants probe  out to  a few kiloparsec.   With a
sufficiently  large sample,  even  apparently bright  stars  will contain  a
statistically relevant sample of halo stars.

The most  efficient use of 6dF is  when exposure times on  one field matches
the set-up time  for the next field (see below) and  this implies a limiting
magnitude of around $I\!=\!12$.  As noted above, RAVE is a precursor to Gaia
and the  wavelength range for the RAVE  spectra was chosen to  match that of
the  Gaia Radial  Velocity Spectrometer  (Munari  2003, Katz  et al.~  2004,
Wilkinson et al.~2005), namely around the Ca II IR triplet.  This wavelength
range  also includes  lines due  to  iron, calcium,  silicon, magnesium  and
titanium, and detailed analyses should provide an estimate of [$\alpha$/Fe],
in  addition  to  overall   metallicity.   Grids  of  synthetic  spectra  at
resolution and wavelength range similar  to those of RAVE, and covering wide
ranges  of $T_{eff}$,  [M/H], $\log  g$, and  $V_{rot}$, were  calculated by
\citet{zwitter04}. and are shown in their Fig.~4.

A typical  RAVE spectrum  is illustrated in  Fig.~\ref{f:nicespectrum}. This
shows a spectrum of HD 154837 (K0~III), as observed on Sep 24 2004, with the
continuum normalized  to 1.0.  The  entire wavelength range is  dominated by
absorption lines;  the strongest due  to FeI, SiI, TiI  and $^{12}$C$^{14}$N
are identified.  The  hump in the continuum around  8508~\AA\ is produced by
an opacity  minimum, due  in particular to  the absence  of $^{12}$C$^{14}$N
lines.

\begin{figure}[hbtp]
\centering
\includegraphics[width=8cm]{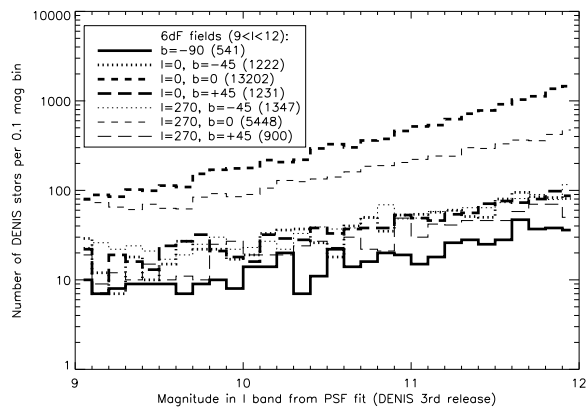}
\caption{DENIS star count distributions as a function of DENIS $I$
magnitude, using the RAVE selection criteria of 9$<$I$<$12 within 6dF fields
of view (5.7 degree diameter),  pointing in the Galactic cardinal directions
visible from the Southern hemisphere.  The total number of stars matching the RAVE criteria are given in parentheses in the  key to the line types. }
\label{f:denis_geo}
\end{figure}

This  wavelength   window  implies  that  an  $I$-band   selection  is  most
appropriate, and  this is indeed the  approach taken.  Star  counts from the
DENIS catalog to a limiting magnitude of $I\!=\!12$ at typical latitudes and
longitudes of RAVE  are shown in Fig.~\ref{f:denis_geo} and  show that a few
set-ups  with 6dF  per line-of-sight,  with a  random selection  over color,
provide a statistically significant sample of the stars on the sky.

\clearpage

\clearpage

\subsection{Description of the instrument}
\label{s:instrument}

The RAVE survey  instrument is known as 6dF \citep{6df}  in reference to its six degree diameter field of view.   
6dF consists of an off-telescope  robotic fiber positioner,
two fiber-field plates  of 150 fibers each and  a bench-mounted spectrograph
which is  fed from  the UK Schmidt  telescope (UKST)  when one of  the field
plates is mounted  on the telescope.  The light received  from the target is
deflected down  the fiber by a  90$^\circ$ prism contained in  a `button' at
the end of the fiber.  This button is magnetic, allowing secure placement of
the  fiber on  the field  plate.  The  robot uses  a  pneumatically actuated
gripper to pick  up, and to put  down, the fiber buttons.  Each  fiber has a
diameter of $100 \mu$m (6.7~arcsec on the sky), and can be placed accurately
(to within 10~microns, or 0.7~arcsec)  on star positions anywhere within the
6-degree diameter  field.  Each field  plate also contains 4  fiducial fiber
bundles  of 7~fibers arranged  in a  hexagonal pattern,  which are  used for
field acquisition.

The bench-mounted Schmidt-type spectrograph sits on an 
optical  bench  on  the  floor  of the  telescope  dome.   The  spectrograph
therefore does  not suffer from  the flexure that  affects telescope-mounted
spectrographs  which are  subject  to  the changing  gravity  vector as  the
telescope moves across the sky. Nevertheless, it is sensitive to temperature
changes  (see  Section~\ref{s:zeropoint}).  With  RAVE  we  use a  specially
purchased  volume phase  holographic  (VPH) transmission  grating of  medium
dispersing power; this 1700 lines/mm grating is tuned for high efficiency in
the I-band, and identifed as grating 1700I.  This setup provides in practice
an  average  resolving power  of  $\sim$7,500  over  the Ca-triplet  region,
covering the wavelength range 8,410 to 8,795~\AA.

The  CCD used  with  the 6dF  spectrograph  is a  Marconi (EEV)  CCD47-10-BI
detector.  It has 13~$\mu$m pixels with a ${\rm 1k \times 1k}$ format and is
thinned  and back-illuminated.  The  actual CCD  dimensions  are ${\rm  1056
\times 1027}$ pixels. The chip has  good cosmetics, with few defects and has
a quantum  efficiency of  40\% to  30\% over the  wavelength region  of RAVE
operations ($\lambda\lambda$8,400--8,800\AA).

The  RAVE  spectrograph  configuration,   with  the  medium  resolution  VPH
transmission grating, unfortunately exacerbates  the effects of the residual 
aberrations  within the Schmidt  system.  This leads  to variable,
position-dependent  PSFs  and  pin-cushion distortions.   Consequently,  the
existing   pipeline  reduction   software,  which   was  designed   for  the
lower-resolution 6dF  galaxy redshift survey,  does not work  optimally with
RAVE  data  (but is  more  than  adequate  for quick-look,  quality  control
purposes).  A dedicated IRAF \footnote{  IRAF is distributed by the National
Optical Astronomy  Observatories, which are  operated by the  Association of
Universities for  Research in  Astronomy, Inc., under  cooperative agreement
with the National Science  Foundation.} pipeline was therefore developed for
the reduction of RAVE data (see Section~4 for details).

Also most  VPH gratings exhibit  a `ghost', due  to light reflected  off the
detector.  This  ghost manifests itself as  a spurious emission  peak in the
spectra and cannot  be avoided in our RAVE  observations.  The wavelength of
this ghost  feature can be  pushed into the  blue part of the  spectra using
hardware tuning, and this wavelength region may then be excluded in the data
analysis e.g.~when computing  the radial velocity  using cross-correlation
techniques.   Not  removing  this  feature  properly  would  result  in  the
correlation function  having a  strongly asymmetric profile.  Fortunately, this
effect is  only significant in a small  fraction of our spectra  and this is
noted by  a quality flag  in the corresponding  entries in the  data release
catalog (see Appendix~A Tables~\ref{t:catalog_desc} and
\ref{t:spec_flag}).

\subsection{Instrument performance}

Each field plate nominally has 150 target fibers, which when undeployed form
a ring  round the  periphery of the  6 degree  field. The fibers  are evenly
spaced  around this  ring,  with the  exception  of two  small  gaps at  the
Northern and Southern  field plate edges. Each gap creates  a small `zone of
avoidance' which  has some (small)  impact on the target  distributions that
can  be  achieved.   Similarly,  there  is  a zone  of  avoidance,  of  less
significance, associated  with the  pivot positions of  each of the  4 guide
fiber  bundles, which  are  located approximately  at  the cardinal  E-W/N-S
points on each field plate.  Each target fiber can nominally reach the field
centre, +10\%,  and its deployed location  relative to parked  is subject to
the constraint  that the angle from  the pure radial direction  must be less
than $\pm 14$ degrees.  The targets  in a given input field are allocated to
a  given fiber using  a sophisticated  Field Configuration  Algorithm (FCA),
based  on that  developed for  the  2dF spectrograph  \citep{fca}.  The  FCA
accepts  user-supplied priorities  within the  input target  list,  but does
contain    subtle    allocation    biases,    as   described    by    Outram
(2004)\footnote{http://www.aao.gov.au/local/www/brent/configure.}        and
\citet{miszalski}.

\begin{figure*}[hbtp]
\centering
\includegraphics[width=7cm]{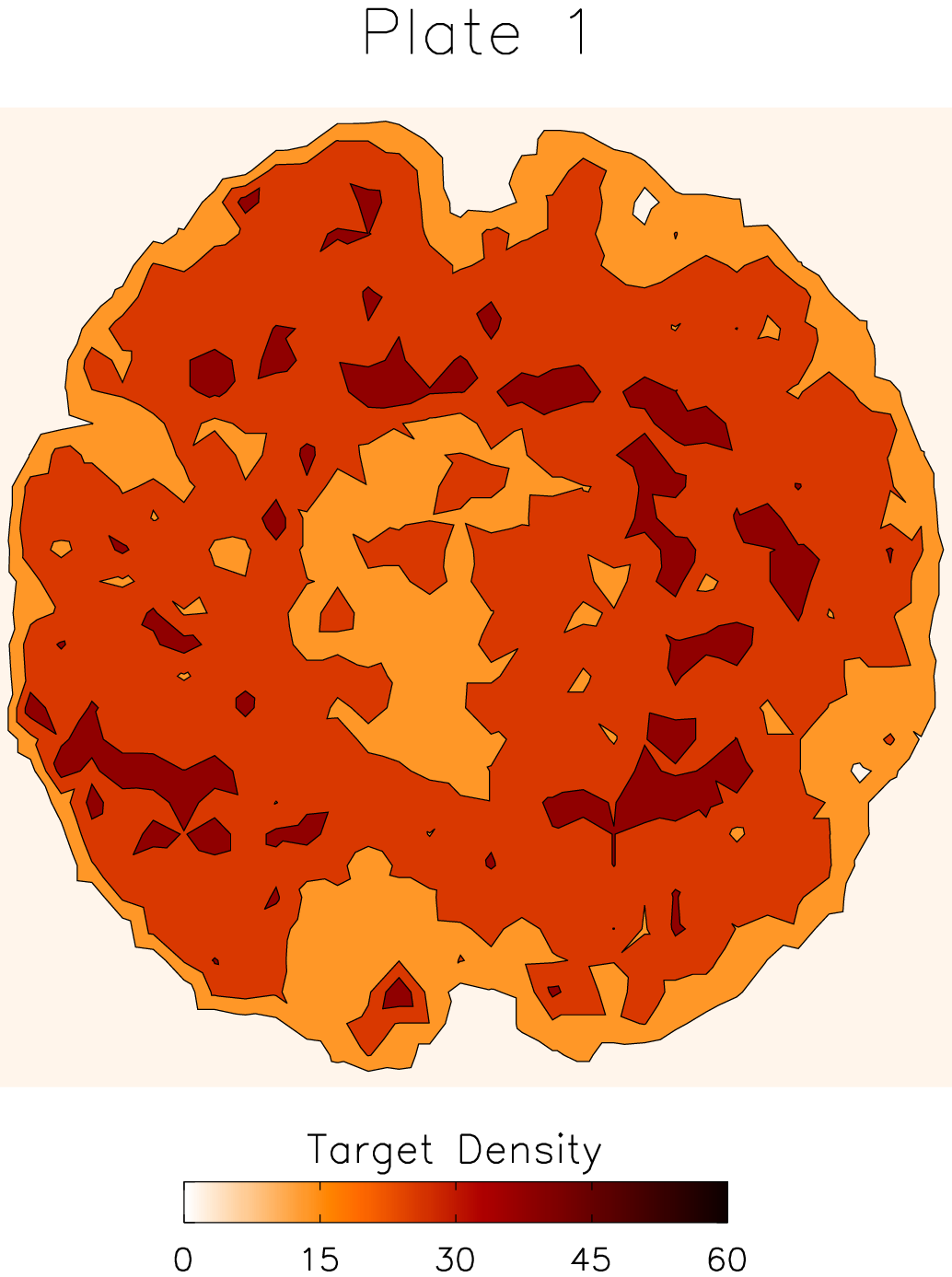}
\includegraphics[width=7cm]{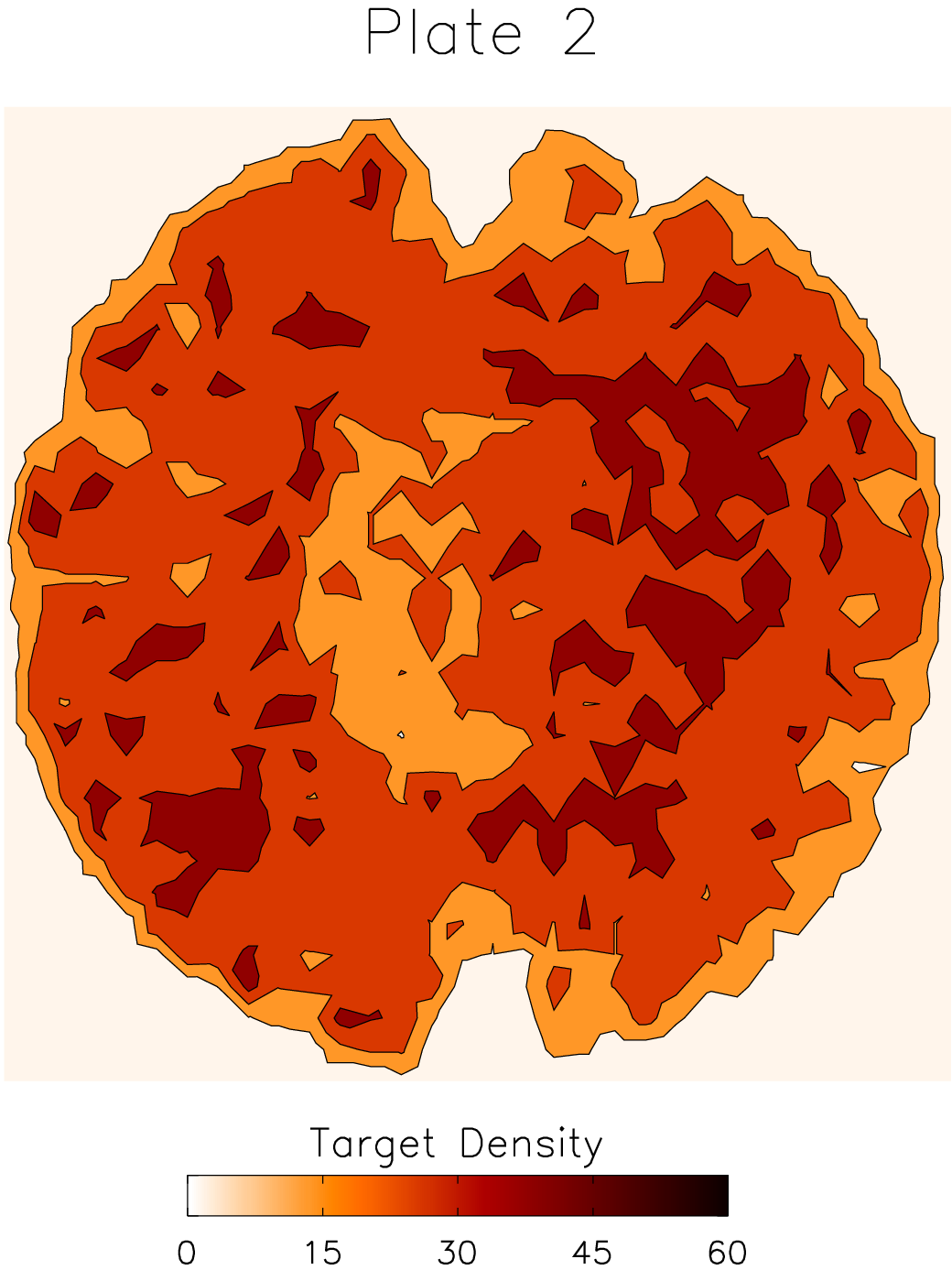}
\caption{The two  panels show the density of  successfully allocated targets
across each field plate from RAVE's first year data. The contours are at 15,
30 and 45 allocated targets per square degree. The structure visible in each
plot is  due to (i) the northern  and southern fiber gap  across each plate,
(ii) zones  of avoidance around each  of the 4 equally  spaced guide fibers,
(iii) known biases within the {\sc CONFIGURE} program (most obvious of which
is the central lefthand under-density) and (iv) broken fibers.}
\label{f:footprint}
\end{figure*}

These  subtle allocation  biases in  the FCA  are illustrated  for  the RAVE
targets  in   Fig.~\ref{f:footprint}  which  shows  contour   plots  of  the
successfully allocated targets, for the first year data, for each of the two
field plates.  The contour levels indicate the number of allocated stars per
square degree.  Since stars are  to first order uniformly distributed across
the  field-of-view,  the   non-uniform  distribution  of  allocated  targets
highlights the  inherent bias in the  fiber placement. The  empty notches at
the  top and  bottom of  each field  plate are  where no  fiber  buttons are
positioned.  The central  under-allocated region is due to  the known biases
in the CONFIGURE program, by which  the central region is considered easy to
reach  and therefore left  until after  harder targets,  close to  the field
edge,  have been  allocated.  Often  this leaves  fewer fibers  than targets
available for  the centre. The remaining  structure in these  figures is the
result  of fibers  being unavailable  (e.g.~due  to breakages)  or having  a
shorter  fiber length  available  (after repair),  re-inforcing the  central
deficiency.\\

\clearpage

\clearpage

Over the current survey lifetime both field plates have had, on average, 130
fibers available  for allocation, an extremely consistent  overall level for
each.  The $\sim$15\%\ of fibers  that are unavailable for allocation at any
one time are due  to a variety of causes.  The most  common problem is fiber
breakages, that are largely the  result of retractor problems on parking the
fiber.   Another problem  is  fractures  in the  fibers  themselves (due  to
buttons `stamping'  on them during  fiber placement, or having  the trailing
short metal  shank of  the fiber  bent on parking,  via collisions  with the
pivot point).  Fractures  may still allow light to  be transmitted, but lead
to severe  fringing effects in  any resultant spectrum.  Another  problem is
throughput deterioration of  the fibers, possibly due to  a gradual clouding
over time  of the UV curing optical  adhesive used to mate  fibers to prisms
(as was the case for the  earlier UKST MOS system FLAIR \citep{flair}).  The
final problem  is related to  the fiber-button gripper mechanism,  which can
lead to the robot being unable to pick-up or put-down some fibers.

Among the problems  listed above, the first two are  linked to the integrity
of the  fibers and  can be solved  by repairing  them.  Fiber repairs  are a
difficult and time-consuming process  which usually involves a shortening of
the  fiber itself  which, after  several repairs,  can reduce  its effective
range.  Nevertheless  considerable effort has been expended  to maintain the
multiplex gain of each field plate.

\subsection{Sample selection and input catalog}
\label{s:input_catalog}

The original RAVE sample was intended to be magnitude-limited over the range
$9\!<\!I\!<\!12$, with no color selection  (where $I$ refers to Cousin $I$).
At the  time of the  initial stages, data  from the 2MASS and  DENIS surveys
were  not yet  available, and  the  RAVE target  stars were  drawn from  the
Tycho-2 catalog  of the 2.5  million brightest stars \citep{tycho2}  and the
SuperCOSMOS  Sky Survey  (Hambly et  al.~2001a, hereafter  SSS).   The Tycho
observations were made in two non-standard filters, $B_\R{T}$ and $V_\R{T}$.
These  passbands   were  converted  to   Johnson  $B$  and  $V$   using  the
transformation (ESA 1997)
\begin{eqnarray}
V = V_\R{T} - 0.090\,(\!B_\R{T}\!-\!V_\R{T}\!) \mbox{,}\nonumber\\
B = V\,\,\, + 0.850\,(\!B_\R{T}\!-\!V_\R{T}\!) \mbox{.}
\end{eqnarray}

$(\!B\!-\!V\!)$  was  trans\-formed  to  $I$ using  an  empirically  derived
zeroth-order mean color transformation  derived from Bessell (1979) Tables~2
and 3:

\begin{eqnarray}
(\!V\!-\!I\!) = 1.007\,(\!B\!-\!V\!) + 0.03 & \mathrm{if}\,\, (\!B\!-\!V\!)<1.30 \mbox{,}\nonumber\\
(\!V\!-\!I\!) = 2.444\,(\!B\!-\!V\!) - 1.84 & \mathrm{otherwise.}
\end{eqnarray}

The  SSS  photometry  \citep{scos2}   is  photographic  $I$  (IVN  emulsion,
hereafter $I_\R{IVN}$), extracted  from the SRC-I UKST IVN  Survey plates by
SuperCOSMOS, the automatic plate scanning facility at the Royal Observatory,
Edinburgh. \citet{photoi}  showed $I_\R{IVN}$  to be directly  equivalent to
Cousin  $I$,  essentially without  any  color  correction,  with a  relation
between the two magnitudes $$I-I_\R{IVN}=0.00\pm0.03 \times (V-I)\mbox{.}$$

The $11\! <\! I\! <\! 12$  interval of the first-year input catalog consists
only  of SSS  stars.   Stars with  \mbox{$9\!<\!I\!<\!11$}  are mainly  from
Tycho-2 but  also includes SSS stars  that do not appear  within 7~arcsec of
any  Tycho-2 star  (this requirement  is set  by the  6dF fiber  diameter of
6.7~arcsec).  Any  SSS star  that does appear  within 7~arcsec of  a Tycho-2
star is not included, and neither is the Tycho-2 star. This procedure allows
us to avoid any contamination by a possible nearby star.  The quality of the
photometry  (not the  astrometry)  in  both catalogs  is  not sufficient  to
discriminate between  the two possibilities that  either (i) the  two are in
fact the same star measured to  have different positions in each catalog, or
(ii) the two are in fact two different stars.  The criteria for inclusion in
the input catalog are designed to merge the two samples, taking into account
both their incompletenesses  at $I\!\sim\!11$ (the faint end  of Tycho-2 and
the bright  end of SSS)  and the better  proper motion accuracy  provided by
Tycho-2.   As a result,  the interim  input catalog  contained $\sim$300,000
stars with $\sim$50\% Tycho-2 and $\sim$50\% SSS.

We  selected  478 contiguous  survey  fields  to  cover $\sim$12,200  square
degrees of  the Southern sky visible  from the UKST,  excluding regions with
Galactic latitude $|b|<14^\circ$, to  minimize dust obscuration and crowding.
The field centres  are defined on a $5.7^\circ$  grid spacing, corresponding
to the field of view of the 6dF field plates.  The gaps in between the fixed
circular  tiling scheme miss  about 20\%  of the  available area.   For each
field of view, 400 targets were  randomly selected from the input catalog to
construct two field files, consisting of 200 stars in each, so that at least
two separate 6dF pointings could be made.

No sub-selection  into bright and faint  samples was made for  the first 2.5
years of  the RAVE  survey, including the  observations for this  first data
release. Occasionally this leads to,  for example, a 9$^{th}$ magnitude star
being adjacent on the slit  to a significantly fainter, 12$^{th}$ magnitude,
star. The tight spacing of the 150 fibers along the slit, as imaged onto the
CCD, means that  about 4\% of each spectrum's flux  contaminates that of the
adjacent  fibers (this effect  is known  as fiber  cross-talk).  This  is an
insignificant  problem  for  radial  velocity determinations  for  the  vast
majority  of  our targets,  but  can  impact  the abundance  determinations,
especially  when  the apparent  magnitude  difference  between two  adjacent
fibers is large.  This cross-talk  effect is carefully taken into account by
the reduction pipeline using iterative cleaning (see Section~4.2).

Typically there are  $\sim$200 potential targets for each  field pointing at
the RAVE magnitude limits  which ensures efficient fiber configurations. The
6dF CONFIGURE software is usually able to allocate all the available science
fibers  to targets, unless  two target  stars are  closer together  than the
minimum allowed separation dictated by  the size of the 6dF button footprint
on  the field-plate  (approximately 5~arcminutes).   For the  first  year of
RAVE, typically $\sim$130 fibers are  allocated to target stars for a single
pointing.   This  offers  scope  for   two  pointings  on  some  fields  and
re-observation of selected targets (repeats).

\begin{figure}[hbtp]
\centering
\includegraphics[height=7.25cm]{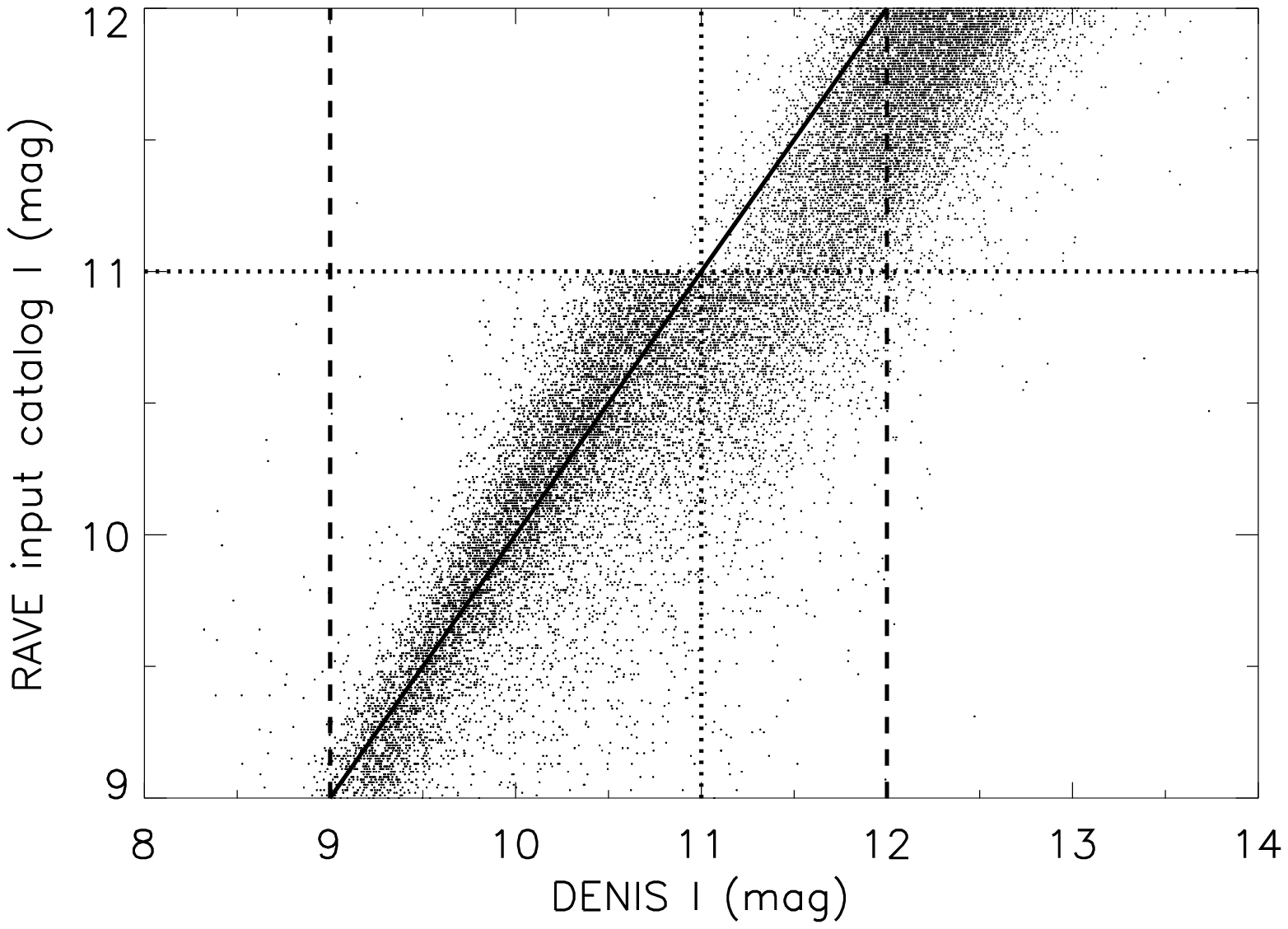} \\
\includegraphics[height=7.25cm]{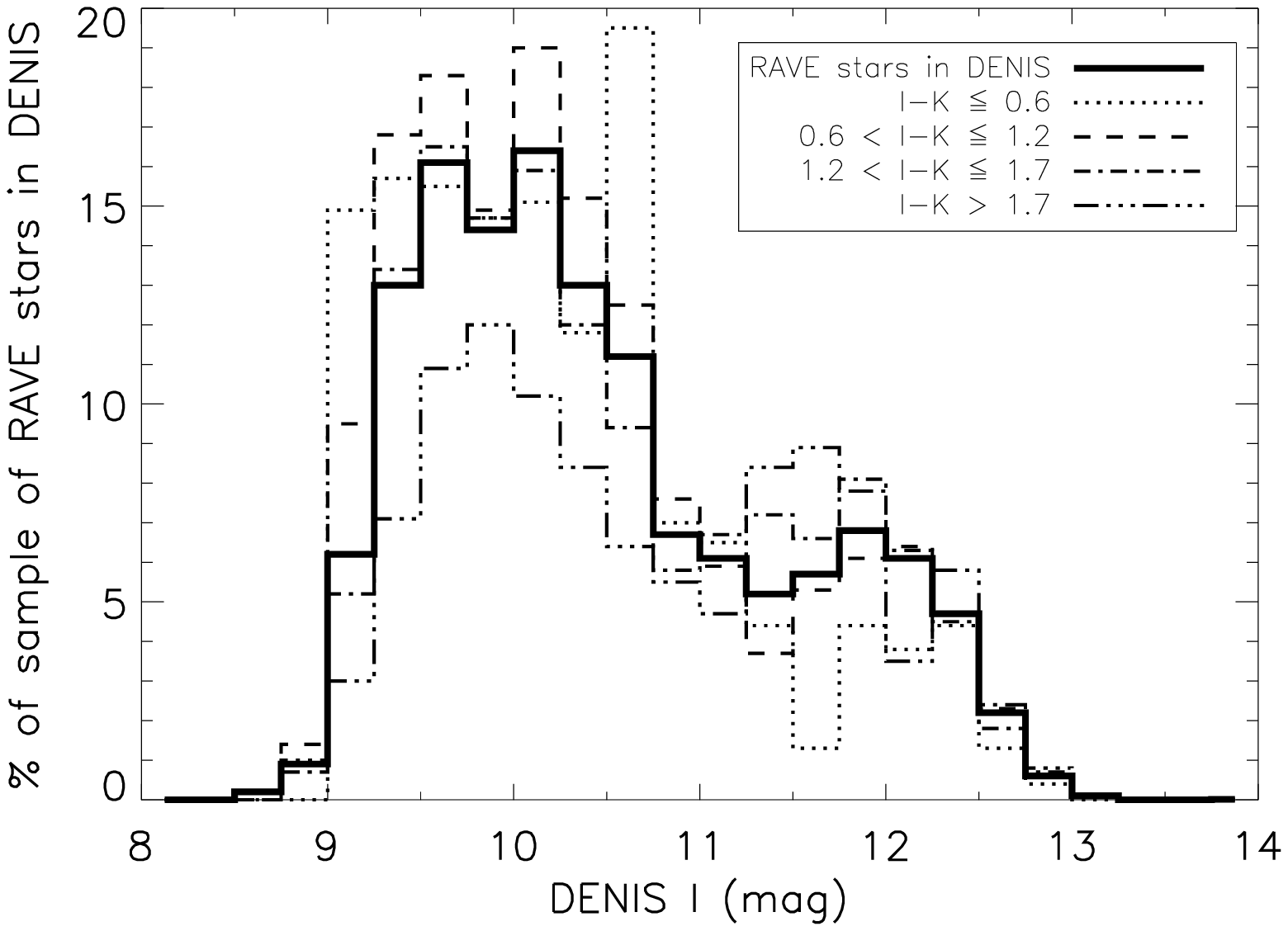}
\caption{Top: Input catalog $I$ magnitudes (Tycho-2 stars entirely within
vertical  axis $I<11$  and SSS  predominately within  vertical  axis $I>11$)
plotted as  a function of DENIS $I$  for a subsample of  observed RAVE stars
that are also in the second  DENIS data release.  The dashed lines represent
the planned  selection window  on each axis.   The dotted line  delimits the
bright  and  the  faint  samples  at  $I$=11.   Bottom:  Histogram  of  RAVE
completeness compared  to DENIS,  as a function  of DENIS $I$  magnitude and
DENIS $I$-$K$ color.}
\label{f:input_cat}
\end{figure}

Both Tycho-2 and SSS are  primarily astrometric catalogs and so only provide
approximate  photometry.  The  second  DENIS data  release,  made public  in
2003\footnote{The latest DENIS release in 2005 is available at the CDS using
the VizieR facility. At the time  of this study, only the second incremental
release  was available.},  presented the  first opportunity  to  compare the
input catalog  $I$ directly with  more accurate $I$-band  photometry (better
than 0.1 mag).  As noted earlier, $I_\R{IVN}$ is equivalent to standard $I$.
\citet{gunni}  showed  that  there  is  essentially  no  difference  between
Gunn-$i$ (DENIS $I$) and $I_\R{IVN}$,  allowing a direct comparison of input
catalog  $I$ with DENIS  $I$.  Fig.~\ref{f:input_cat}  (top) shows  that the
\mbox{Tycho-2} $I$--magnitudes derived from the \mbox{Tycho-2} $B_\R{T}$ and
$V_\R{T}$  photometry systematically agree  with DENIS  $I$ (mean  offset $=
-0.095$~mag), albeit with  a large scatter ($\sigma =  0.385$~mag).  The SSS
$I$--magnitudes appear  to be  nonlinear, systematically diverging  from the
DENIS $I$ with brightness (mean  offset $= -0.50$~mag, $\sigma = 0.33$~mag).
DENIS provides the first large-scale,  external check on the accuracy of the
SSS photometry  for bright  stars.  This diverging  offset reflects  the SSS
zeropoint error in linearizing the  non-linear saturation of bright stars on
photographic plates.   A result is that  the number density of  SSS stars in
our sample,  shown in  Fig.~\ref{f:input_cat} (top), actually  peaks fainter
than the planned selection window.

Fig.~\ref{f:input_cat} (bottom) compares a  subsample of observed RAVE stars
with the  number of DENIS stars  in the same area  of sky, as  a function of
DENIS $I$ magnitude.   The black line emphasizes that  RAVE is not complete.
The colored lines highlight color  biases present in the input catalog.  The
bright sample  ($I<11$, mainly Tycho-2)  has relatively more blue  stars and
fewer red stars  than has the faint sample ($I>11$,  entirely SSS), and vice
versa,  reflecting the  effective selection  biases in  the  two subsamples.
Although  the  pseudo \mbox{$I$-band}  selection  from  Tycho-2 has  reduced
\mbox{Tycho-2's}  $B$-selection   bias,  the   bias  is  still   visible  in
Fig.~\ref{f:input_cat},  compared  to  real  $I$-band  selection  from  SSS.
Hence,  the   input  catalog  is   not  an  homogeneous   selection  window:
sample-dependent  color  biases  exist.   In  terms  of  stellar  population
studies, one of RAVE's scientific goals is to target red giants, ($1.2 < I -
K \le 1.7$),  to probe the Galactic  disk and halo.  This goal  has not been
compromised by the photometry on which the interim input catalog was based.

To summarize, the catalog has  no kinematic bias, but Galactic science using
RAVE requires care to account for the various selection biases introduced by
the  inhomogeneous photometry  used to  derive  the input  sample.  We  must
stress that the RAVE catalog in the first data release is clearly incomplete
within its selection criteria (and was  not intended to be complete) and has
subtle color biases.  The distribution  of RAVE stars in the color-magnitude
diagram  is   not  entirely   representative  of  the   underlying  Galactic
population, due  to the selection effects described  above.  Subsequent data
releases  will  include  stars  selected  by $I$-magnitude  from  the  DENIS
catalog, which should provide  an essentially unbiased representative sample
of Galactic stars in the selected range of magnitude.

\clearpage

\begin{figure}[hbtp]
\centering
\includegraphics[width=7.5cm]{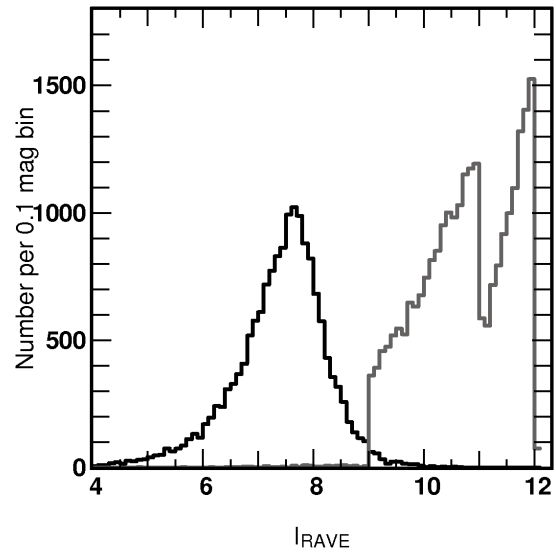}
\caption{Comparison  of  RAVE  (grey)  and  GCS  (black)  apparent  magnitude
distribution, showing  that RAVE is significantly fainter  than the previous
large radial  velocity survey.   The GCS data  (Nordstrom et  al.~2004) have
been transformed to RAVE's input catalog $I$-band magnitude using Eqs.~1 and
2 of the  text.}
\label{f:Imag_geneva_rave}
\end{figure}

\clearpage

Fig.~\ref{f:Imag_geneva_rave}   compares  the   $I$-band   apparent  magnitude
distributions  of the  RAVE sample  and for  the Hipparcos-based  GCS sample
(Nordstrom et al.~2004), which is the largest previously published sample of
accurate  stellar  radial  velocities.   The  RAVE  $I$-magnitudes  for  the
brighter RAVE stars  and the GCS sample are derived from  B and V magnitudes
using equations 1  and 2 above.  The RAVE and GCS  samples hardly overlap in
apparent  brightness.  The  RAVE magnitude  range ($9\!\!<\!\!I\!\!<\!\!12$)
was  chosen to  match  the scientific  goals  of the  RAVE  survey with  the
instrumental  capabilities  of  the  6dF  system.  Therefore,  as  shown  in
Fig.~\ref{f:Imag_geneva_rave}, the RAVE sample  is much fainter than the GCS
sample. This will be relevant when we compare the RAVE velocities with those
from external datasets (section~\ref{s:external_datasets}).

\subsubsection{RAVE Catalog contamination}

Although RAVE target samples are ostensibly single stars selected from Tycho
and SSS-$I$,  both samples,  including Tycho, do  in fact  contain double-star
contaminants. This  was noticed during the  early part of the  survey from a
casual  examination  of  printed   target  images.   There  are  three  main
situations which arise: a) Close  doubles of similar magnitude where the 6dF
fiber will be  positioned on the centre of the double  image and a composite
spectrum will  result; b) Close pairs  where one star is  very much brighter
than the other.  The position  for these `binaries' will be heavily weighted
to  the  bright  component and  so  only  a  small  effect will  result;  c)
Relatively close pairs  (and this is a situation which  only arises with the
SSS data at the fainter end) where a blended SSS image results but where the
SSS position is between the two stars in the blend.  In such cases the fiber
would   be  positioned   essentially  on   sky.   An   additional   case  of
multiple-blends of  3 or more stars  is so rare  that it can be  ignored. To
assess the extent  of the problem, 200 randomly selected  fields from year 1
data were chosen  and each `thumbnail' small-area image  for each target was
carefully  examined.  This  established that  the problem  was present  at a
maximum  of 0.4\%  level. As  a further  check a  blind comparison  of these
contaminants was made with the associated spectra to see if such blends were
obvious.  Only  a fraction (25\%) of  the problem stars were  obvious in the
spectra.   Now we  have  a rigorous  regime  where all  candidate stars  are
checked for  possible contamination prior  to observing, using  the 1-arcmin
SSS  thumbnails from  the on-line  SSS $R$-band  data.  Such  contaminants are
identified and removed from the database.

\section{Observations}
\label{s:observations}

Observing time during  the first year of the RAVE survey  was confined to an
average of  seven bright/grey nights  per lunation, distributed  around (but
not including)  four or five unscheduled bright-of-moon  nights. The project
began on 2003 April 11, and  the first year's data includes spectra obtained
between then and 2004 April 3, for a total of 88 scheduled nights.

Observations for  RAVE consist of a  sequence of target  field exposure, arc
and flat. During the first year, Ne, Rb and Hg-Cd calibration exposures were
obtained  for each field,  together with  a quartz  flat field  for spectrum
extraction in the data reduction. All calibration exposures were obtained by
reflection from a white full-aperture  diffusing screen located close to the
entrance pupil of the telescope  (the Schmidt corrector).  In this position,
the  effects of  irregularities in  the screen  illumination  are minimised,
while the fibers are illuminated at a similar focal ratio (f/2.5) to that of
the incoming starlight. The requirement  for arc and flat observations to be
obtained for  each field arises because  the slit units  are interchanged in
the spectrograph for each new  configuration (along with the field plates in
the telescope).

Several  target  fibers are  reserved  to  monitor  the sky  for  background
subtraction.  As RAVE  stars are quite bright, only a  small fraction of the
signal in  each fiber comes from the  sky.  In most cases  only sky emission
lines are present, with little  or no background of scattered solar spectrum
from  the Moon,  despite the  fact that  most first  year  observations were
obtained in bright or grey time.   It is thus necessary to subtract this sky
using measurements  from dedicated sky  fibers, placed uniformly  across the
field. A calculation of the optimal  number of sky fibers per plate is given
in   Kuijken  \&   Gilmore  (1989b),   while  a   detailed   description  of
sky-subtraction with fibers is given in Wyse \& Gilmore (1992).  Each of the
RAVE target frames therefore contained  spectra of at least ten sky samples,
obtained using dedicated  sky fibers. These were combined  and scaled in the
reduction process for sky subtraction.

On a subset  of nights (of order one in four),  twilight exposures were also
taken to provide additional zero-point velocity checks from solar absorption
features and twilight sodium D lines. These were obtained with the telescope
pointing towards  the zenith, and taken  within thirty minutes  of sunset or
sunrise. Those twilights  frames are particularly useful to  assess the zero
point stability and enable some more radial velocity checks.

The target  frames themselves consist  of five consecutive exposures  of 600
seconds each, allowing adequate signal-to-noise to be obtained in the summed
spectra, while  minimizing the risk  of saturation from  particularly bright
stars. In poor  conditions of low sky transparency, additional 
exposures are  made. The total  time for a  `standard' set of arc,  flat and
field  exposures  is of  order  one  hour, which  is  similar  to the  fiber
reconfiguration time in the 6dF  robot.  Thus, while observing proceeds with
one  field   plate,  the  other   is  being  reconfigured,   maximizing  the
time-on-target  productivity  of  the  telescope.   Taking  account  of  the
physical transportation and exchange of  the field plates, the slew time for
the telescope,  field acquisition etc.,  an experienced observer is  able to
accumulate acceptable  data for  up to eight  RAVE fields on  a mid-winter's
night at the latitude of Siding Spring Observatory.

The median seeing at the focal surface of the Schmidt Telescope
(i.e. including local site effects and convolved with the
instrumental PSF) is $\sim$2.5 arcsec. This is much less than the
6.7 arcsec diameter of the fibres, and so some seeing degradation
is possible without serious loss of flux. In extremely poor
conditions, however (seeing $\sim$4 arcsec), there is a noticeable
loss of signal.  This is principally due to the difficulty in
obtaining perfect target acquisition because the seeing degrades
spatial resolution in the guide fibre bundles. Fortunately, such
poor seeing is relatively rare ($\sim$10--15 percent of clear
time).

During the  first year of  RAVE operations, no  blocking filter was  used to
isolate the spectral range required,  so there is second order contamination
of the  spectra. Flux in  the approximate wavelength  range $\lambda\lambda$
4200--4400  \AA~is therefore  added to  that in  the primary  spectral range
($\lambda\lambda$  8410--8790 \AA).   In practice,  this is  only  a serious
problem for hotter stars (earlier spectral types than $\sim$F), but features
such as H$\gamma$ will be present at a low level in all stars.

The final tally for the first year's observations was 24,748 target stars in
240 fields.   Of these stars,  24,320 were observed  once, 330 twice  and 98
three times.  A  small subset of these stars  ($\sim$100) were also observed
with the  2.3-m telescope  of the Australian  National University  at Siding
Spring Observatory and the ELODIE \'echelle spectrograph at the Observatoire
de Haute-Provence (France), and a further  small number was found to be part
of  the GCS  sample. Those  observations are  described in  more  details in
section~\ref{s:external_datasets}.

\section{Data reduction}
\label{s:data_reduction_main}

Data  reduction  for the  RAVE  project consists  of  two  phases.  First  a
quick-look  data  reduction  is  carried  out at  the  telescope  using  AAO
software.   This was  originally the  2dFdr  package developed  for the  2dF
Galaxy Redshift  Survey.  It  was modified  to reduce 6dF  data for  the 6dF
Galaxy Survey (therefore  named 6dFdr for 6dF data  reduction software).  At
RAVE's high dispersion,  the wavelength modelling of 6dFdr  was too crude to
allow  arc lines to  be correctly  identified.  Therefore,  the spectrograph
optics model in 6dFdr has been upgraded to allow correct line identification
at all dispersions.

6dFdr was used  to reduce RAVE data using simple  fiber extractions from the
raw data frames  (summing pixel values over a fixed  width around each fiber
trace), without subtracting scattered light. 6dFdr was upgraded to RAVEdr in
November 2004 to include scattered light subtraction.  Note that RAVEdr (and
6dFdr) are able  to perform more sophisticated fiber  extractions, which fit
profiles to  each fiber cross-section to reduce  fiber cross-talk.  However,
this critically depends on the accurate subtraction of scattered light.  The
current  scattered  light  model  is  only  valid for  all  stars  of  equal
brightness, which  is not  always the case  for RAVE's  observations.  Also,
RAVE's  bright   star  observations  cause  the   more  sophisticated  fiber
extraction to  be unstable  so the  simple method is  used. This  results in
fiber  cross-talk being not  optimally modelled  with RAVEdr.   Therefore an
IRAF-based  data   reduction  pipeline  has   been  built  to   reduce  RAVE
observations.  This pipeline  is described in section~\ref{s:raw_reduction}.
Nevertheless, the RAVEdr  reductions by the UKST observers  remain useful as
the first of many data quality checks.

\subsection{Raw data reduction}
\label{s:raw_reduction}

Raw data  are reduced with a  custom IRAF pipeline. The  pipeline uses three
types of exposures  to reduce every field: a set  of scientific exposures is
followed by a Neon arc and a flatfield exposure. The flatfield frame is used
to identify the  location on the CCD of the  different spectra of individual
fibers. This  is done  automatically, using the  information that  there are
small additional  gaps following  the fibers with  numbers 50 and  100.  The
pixels  associated  with  all   fibers  exceeding  a  threshold  signal  are
extracted.   This allows the  evaluation of  fiber cross-talk  and scattered
light for any fiber with contributing  light, even if it has been identified
as broken.

The fibers  are relatively close  to each other  on the CCD, with  a typical
fiber-to-fiber  separation  being  approximately  twice the  FWHM  of  their
spatial PSF  profile. Fiber  cross-talk therefore needs  to be  analysed and
included in  the reduction procedures.  We adopted  the following procedure:
first, all light within $\pm 0.7$~FWHM of a given fiber center is attributed
to  this fiber.   Next,  a  Gaussian, modelling  the  contribution from  the
neighboring fibers, is subtracted, and  then the whole procedure is iterated
twice.  We  estimate, from our experience,  that in the  final spectrum less
than 1\% of the flux is due to stars in adjacent fibers along the slit, even
if  those adjacent  stars  are  2.5~mag brighter.   This  level of  residual
contamination is certainly acceptable.

The scattered light background requires a different careful treatment. It is
relatively strong and there are no  large gaps in the distribution of fibers
on the entrance  slit that allow for its  easy determination.  The scattered
light is  modelled by spreading a fraction  of the total stellar  light to a
wide,  axially symmetric  Gaussian profile  (see  Wyse \&  Gilmore 1992  for
discussion). Its width  is $\sim$1/3 of the CCD chip  size and its intensity
$\sim$13\% of the  stellar flux.  These two parameter  values are determined
iteratively by checking the residuals in flatfield and scientific exposures.
Spectra of sky fibers (background) typically feature only sky emission lines
($\sim$90\%  of the  cases), so  their  zero continuum  value is  used as  a
further  check on  the  consistency of  the  scattered light  model. In  the
remaining 10\%, scattered  light from cirrus clouds and/or  moonlight add an
absorption spectrum to the sky  spectra (resulting in a non-zero continuum).
Those  cases are  easily detected  via the  existence of  Calcium absorption
lines  in the  spectra. Here  we  check that  the Ca  lines have  meaningful
intensities (eg.  counts at the center are non negative).

Other tasks performed by  the IRAF pipeline include flatfielding, wavelength
calibration,  sky subtraction and  heliocentric correction.  Flatfielding is
crucial,  as fringing is  very strong  at the  wavelengths of  RAVE spectra.
Wavelength calibration is achieved by a low order polynomial fit to 9 Ne arc
lines. The sky  to be subtracted from all spectra is  determined as a median
of  those  sky  fiber spectra  which  are  found  to  be free  from  stellar
contamination, fiber cross-talk or ghost peaks.

The observed  wavelengths are corrected for  the rotation of  the Earth, the
motion of  the Earth about the  Earth-Moon barycenter, and the  orbit of the
barycenter about the  Sun. The corresponding formulae are  given in the help
file of the procedure {\it rvcorrect} in IRAF.  The resulting transformation
from observed to heliocentric velocities should be accurate to 0.005~km/s.

A  median combination of  a sequence  of five  600~s exposures  produces the
final spectrum with  improved signal to noise ratio (SNR) and essentially free  from cosmic ray hits
(except  for cosmic  ray hits  in  the 250~s  exposure of  the flat  field).
Values of all  parameters used during reduction are  recorded in the reduced
file's  header,  while  any  peculiarities  are  noted  in  the  appropriate
reduction logs.

The procedure  is automated, but the  user is asked to  confirm manually the
crucial steps: fiber identification, scattered light subtraction parameters,
quality of wavelength  solution for each fiber and  the appearance of fibers
with sky (background) spectra. This ensures that errors due to problems with
the weather or the instrument do not compromise the results.  An experienced
user needs $\sim$20~minutes to reduce the sequence of 5 scientific exposures
of a given field, together with their flatfield and arc exposures.

\subsection{Radial velocity}
\label{s:radial_velocity_main}

\subsubsection{Rest frame radial velocity measurement}
\label{s:radial_velocity}

\begin{figure*}[hbtp]
\centering
\includegraphics[width=13cm]{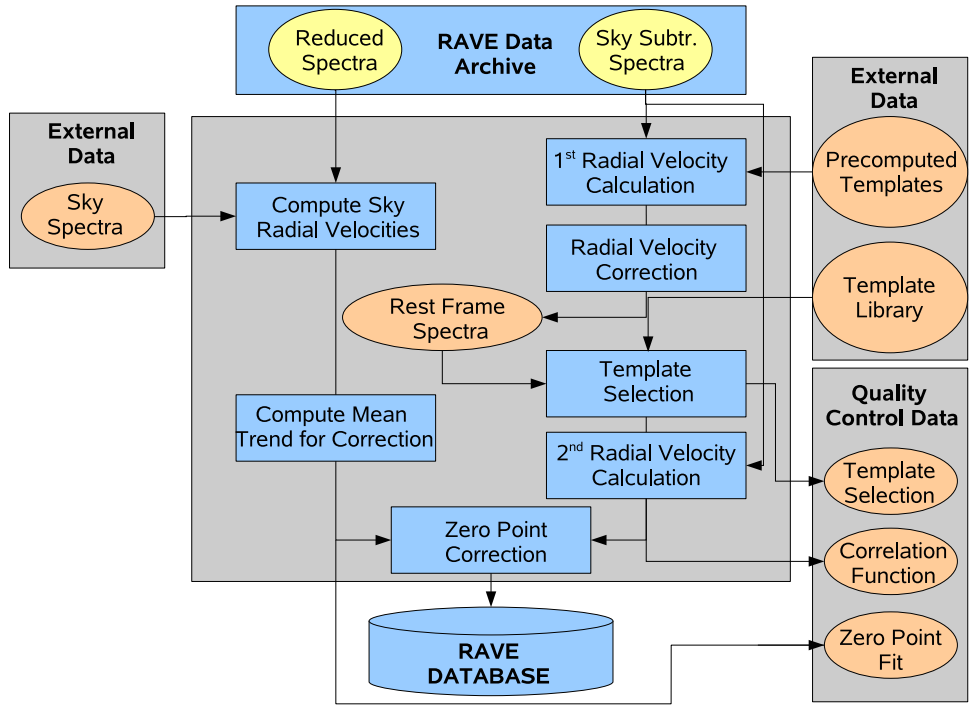}
\caption{Description of  the Radial Velocity Pipeline. The  pipeline is made
  of two  main branches, the  first one (on  the right) computes  the radial
  velocities  from sky-subtracted spectra  while the  second branch  (on the
  left) uses sky unsubtracted spectra to compute the zero-point correction.}
\label{f:RV_pipeline}
\end{figure*}

The measurement of  radial velocities is performed by  an automatic pipeline
(see  Fig.~\ref{f:RV_pipeline} for  a  schematic description)  which uses  a
standard  cross-correlation  procedure  \citep{tonry}.   The speed  of  such
routines,  based on  Fourier transforms,  makes them  well adapted  to large
datasets  like  RAVE where  many  radial  velocity  derivations have  to  be
performed. Here, the pipeline uses the package XCSAO for IRAF \citep{xcsao},
as  it combines  both  the  speed of  the  cross-correlation techniques  and
adequate formats for the outputs.

\clearpage

\clearpage

When computing the  radial velocities, the blue and red  ends of the spectra
are  rejected, the effective  wavelength interval  used being  8,460~\AA~ to
8,746~\AA. The reason for this is to avoid the poorer focus, lower resolving
power and  lower quality of the  wavelength calibration at the  edges of the
spectral interval, while also excluding,  in most of the cases, the emission
ghost    feature   at    $\sim   8450$~\AA\    which   was    discussed   in
Section~\ref{s:instrument}.

Prior to the  radial velocity determination, the spectra  are also continuum
normalized using a  cubic spline.  A cosine-bell filter  is used to minimize
the effects of (high frequency)  noise and the radial velocity is calculated
by fitting a  parabola to the top  of the correlation peak (the  top 20\% of
the correlation function is used throughout the pipeline).\\

Measuring radial velocities using cross-correlation techniques relies on the
availability of  accurate stellar templates  to correlate with  the observed
spectra.  Generally, one  template or a reduced set of  templates is used so
as  to  match a  given  spectral  type\footnote{For  example, the  automatic
reduction pipeline for the  high resolution \'echelle-spectrograph ELODIE at
the Observatoire  de Haute-Provence (used mainly for  an extra-solar planet
search) uses two  templates, matching G and K dwarfs.}.   As the RAVE sample
covers  the entire  color-magnitude  diagram, and  therefore includes  large
ranges  of  stellar  physical  properties  such as  $T_\R{eff}$,  $\log  g$,
$\R{[M/H]}$  and rotational  velocity,  a  large number  of templates  is 
needed  in  order to
obtain accurate  radial velocities for all  targets.  We use  the library of
theoretical  spectra  from \citet{zwitter04}  covering  the RAVE  wavelength
interval.~ This  library contains 62,659 synthetic spectra  at the RAVE/Gaia
resolution,  constructed using  Kurucz  model atmosphere  code (ATLAS).   It
covers   an  extensive  range   of  physical   conditions  of   the  stellar
atmosphere.\\

As  no blue  blocking  filter (OG531)  was  used during  the  first year  of
observations,  the resulting second  order contamination  needs to  be taken
into account. The second order templates (requiring a bluer wavelength range
than  the  Zwitter  et  al.~spectra)  are taken  from  the  \citet{munari05}
library,  which contains  28,180  synthetic ATLAS  spectra  which have  been
degraded to RAVE's resolution. The  combination of both libraries leads to a
set of 22,992  synthetic spectra that are used for  this first release.  The
range  in  effective  temperature  of  this  library  is  $3,500  \R{K}  \le
T_{\R{eff}} \le 40,000  \R{K}$, giving good coverage of  all spectral types.
$\log g$  varies from $0$  to $5$ in  steps of $0.5$, while  the metallicity
[M/H] is  computed for $[-2.5 ,  -2 , -1.5 ,  -1 , -0.5 ,  -0.2 , 0  , 0.2 ,
0.5]$, all scaled solar elemental abundances.  The rotation velocities range
from 0~km/s to 500~km/s, with  irregular spacing.  For the entire library of
synthetic spectra, the microturbulent velocity  is assumed to have the value
of $2$~km/s. \\

The number of template spectra is sufficiently large that it is not possible
to compute the correlation function  for all of them.  Therefore, the radial
velocities are obtained using a  four-step process.  In step~1, we perform a
first  guess of  the radial  velocity using  a reduced  set of  40 templates
covering evenly  the parameter  space of $T_\R{eff}$,  $\log g$  and 1st/2nd
order ratio  (matching spectral type in  the template is  more important for
radial  velocity determination  than is  metallicity).  This  first estimate
provides  a  median  internal  accuracy  of $\sim$5~km/s  for  most  of  the
CaII-dominated spectra.  For  early type stars, where the  Paschen lines are
dominant, the  accuracy is lower, due  to the large width  of those hydrogen
lines.  As a comparison, the pixel size in velocity space for a typical RAVE
spectrum varies  from 12~km/s to  14~km/s from the  red to the blue,  with a
mean value  of 13~km/s.  Therefore, with  careful analysis and  high SNR, we
can  expect to  reach  a  $\sim$1.3~km/s internal  accuracy  for the  radial
velocities (one tenth of a pixel in velocity space).

In step~2,  using this first estimate  of the radial  velocity, the observed
spectrum is  shifted to  the rest  frame and compared  to the  full template
database to select the  `best-matching' synthetic spectra. The best-matching
template  is  defined  as  the  spectrum (a  combination  of  one  1st-order
synthetic spectrum and one 2nd-order synthetic spectrum) which minimizes the
quantity:
\begin{equation}
D=\sum_{\lambda} [O(\lambda)-\left((1-c) S_{1}(\lambda)+c S_{2}(\lambda)\right)]^2\mbox{,}
\end{equation}
where  $O(\lambda)$ is  the continuum-normalized  observed  target spectrum,
$S_{1,2}$ are the  continuum-normalized 1st and 2nd order  templates and $c$
is   the  fractional   2nd  order   contamination.

\begin{figure}[hbtp]
\centering
\includegraphics[width=7cm]{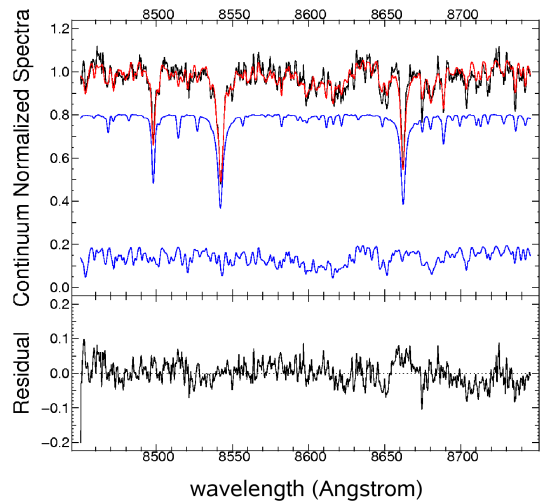}

\caption{An  example of the  quality  of the  template fitting  that is
  produced        by         the        algorithm        presented        in
  Section~\ref{s:radial_velocity_main}, applied to the spectrum of a typical
  target  star, here  T6646\_01424\_1.  The  top panel  shows  the observed,
  continuum  normalized, spectrum  (in  black), together  with the  selected
  template  (red).  The  individual  1st- and  2nd-order  components of  the
  template  are  shown in  blue.  The  second  order contribution  for  this
  particular target  amounts to  less than $\sim  20$~\%.  The  bottom panel
  shows  the residuals  after subtracting  the synthetic  template  from the
  observed spectrum. }
\label{f:template_fit}
\end{figure}

This equation does not account for a color term which is expected because of
the  shape  of  the  continuum  between  the  first-order  and  second-order
wavelength range.   Nevertheless, this effect  is small (due to  the limited
range in  wavelength of the RAVE  spectra, only $\sim  350$~\AA) compared to
other sources  of error, such as  noise in the observed  spectra or residual
scattered light.  Furthermore, including a color term properly would require
continuum normalization  for each  possible combination of  $S_{1}$, $S_{2}$
and  $c$, which is  beyond the  computational time  limit for  the pipeline.
Finally, to speed up selection of the best template, $c$ is not treated as a
continuous variable, but rather is  limited to discrete values between 0 and
0.8,  spaced by 0.1,  covering the  possible values  of this  parameter.  An
example of the  acceptable fit that is the  outcome of this template-fitting
procedure is given in Fig.~\ref{f:template_fit}.  The estimated contribution
of 2nd-order  light to  the observed spectrum  is indicated, plus  the small
residuals after the template is subtracted from the observed spectrum.

At  step~3,  the appropriate  template  having  been  chosen, a  new  radial
velocity is calculated.  In step~4,  this new determination is corrected for
a possible zero-point offset. This  final correction, as well as its origin,
are discussed in the next section.\\

At each of the important steps in this process the pipeline produces summary
plots which  are used to  detect possible problems.  Information  from those
quality-check  outputs are  reported in  the final  catalog released  to the
public. Also, note that the output  spectra of the reduction pipeline are in
the  heliocentric reference  frame  (see section~\ref{s:raw_reduction})  and
therefore, the radial velocities provided in the catalog are heliocentric.

\subsubsection{Zero-point offset origin and correction}
\label{s:zero_point}

Analysis  of raw  arc frames  taken before  and after  observations  of RAVE
fields showed  that the positions of the  emission lines in one  of the arcs
are sometimes  shifted along the  spectral axis on  the CCD relative  to the
other  arc.  The  origin of  this  shift is  most likely  to be  temperature
variations in the spectrograph room, which induce a slight change in some of
the optical  components.  This results in  a small offset  in the dispersion
direction, of  the order of  a few  tenths of a  pixel, between the  two arc
frames. This  will also cause an  offset between a scientific  frame and the
arc frame used to calibrate that  science frame.  The result is a zero-point
offset in the measured radial velocities.  A tenth of a pixel corresponds to
less than 1.5~km/s, and this is then the order of the zero-point offset.

\begin{figure}[hbtp]
\centering
\includegraphics[height=7.5cm,angle=270]{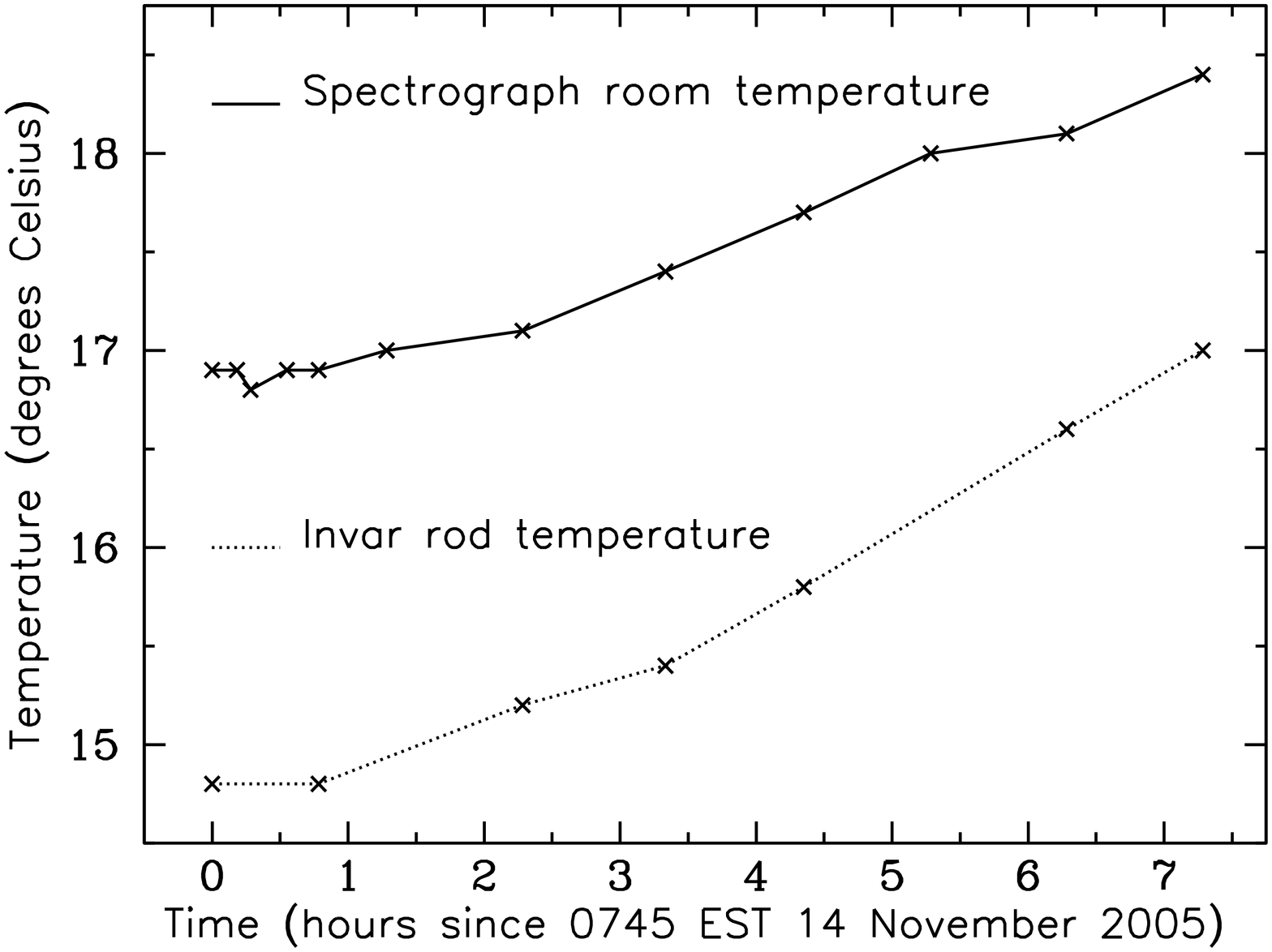}
\includegraphics[height=7.5cm,angle=270]{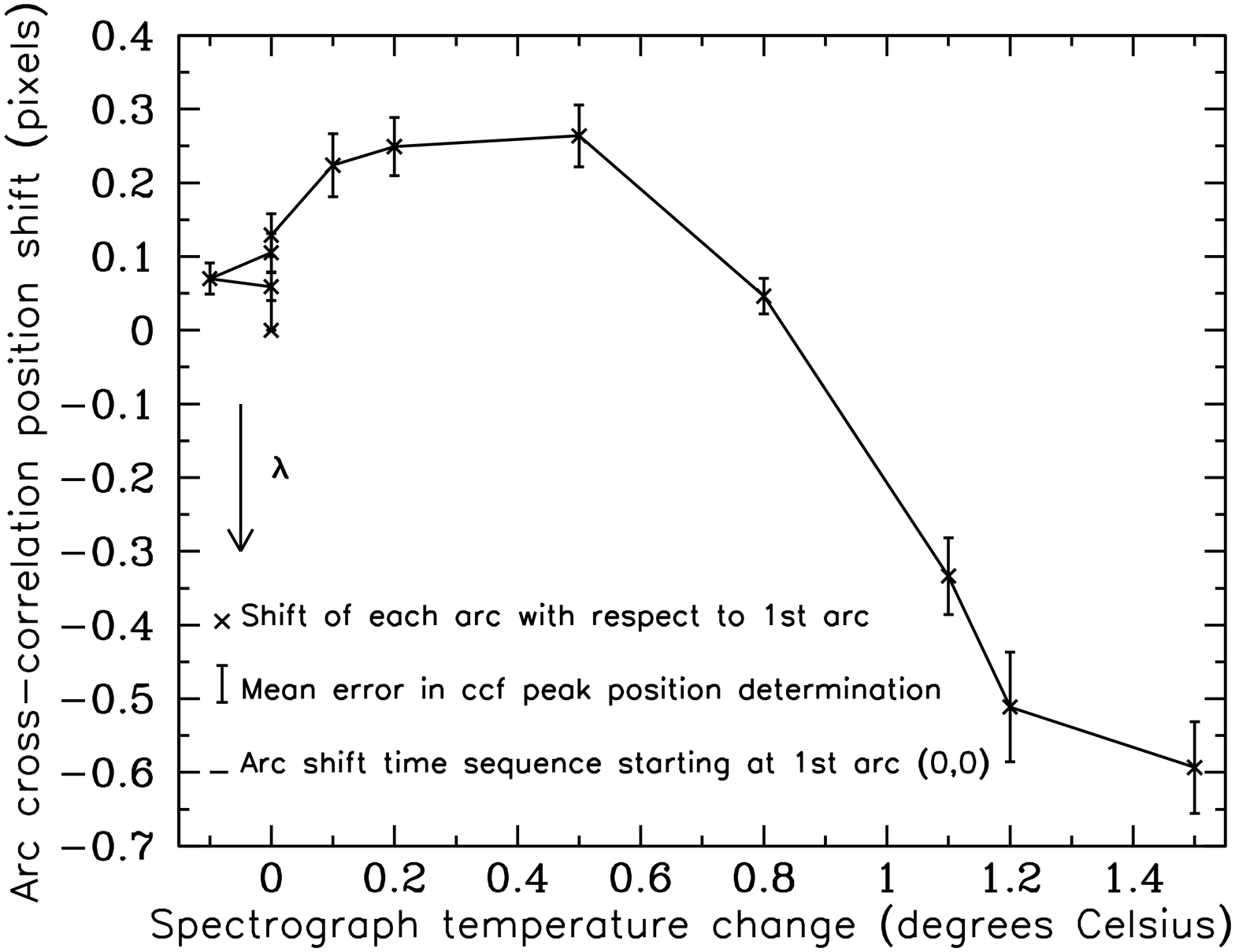}
\caption{Top: Spectrograph room  and invar rod temperatures as  a function of
  time.  Bottom: Shift, as a function of temperature, in the position of the
  cross-correlation peak for  the central arc in one  frame, relative to the
  central arc in a reference frame (the `1st arc').  }
\label{f:pixel_shift}
\end{figure}

Both hardware and  software were investigated to find the  source of the arc
shifts.   The  slit-vane assembly  was  found  to  be vibrationally  stable,
despite sharing the  same mounting on the optical  bench as the spectrograph
shutter.  The spectrograph room temperature is only recorded at the start of
a  RAVE  setup,  so  the  temperature  change over  the  course  of  a  RAVE
observation is not known.   Fig.~\ref{f:pixel_shift} shows the results of an
experiment  to test the hypothesis  that temperature  variations are  indeed the
source of the shifts.  Arc exposures were obtained over the course of a day,
with the  spectrograph room door shut  and the dome  shut.  The spectrograph
room temperature was recorded at the  start of each arc exposure.  The invar
rod  temperature   was  also  recorded   to  check  the   spectrograph  room
thermometer.  (The  invar rod is  within the telescope, between  the primary
mirror and field  plate holder.  Its temperature is used as  an input into a
model to  maintain focus  between the primary  mirror and the  field plate.)
The top of Fig.~\ref{f:pixel_shift}  shows the spectrograph room thermometer
is consistent with the invar rod thermometer.

The central fiber of each  arc frame was extracted and cross-correlated with
the central  fiber of the  first arc frame.  The  cross-correlation function
(ccf) peak position corresponds to the overall emission line shift in pixels
between each  set of arcs.   The bottom of Fig.~\ref{f:pixel_shift}  shows a
negative pixel  shift trend  as a result  of a positive  temperature change,
suggesting that the spectrograph room  is not a thermally stable environment
and that this is responsible for the radial velocity zero-point offset.

If the  glass VPH grating were  expanding with temperature,  the pixel shift
would be positive  according to the grating equation,  implying that this is
not the source of the zero-point shift. The thermal expansion coefficient of
steel is 1.44 times that  of ordinary glass but steel's thermal conductivity
is almost  63 times greater than  that of glass. This  suggests that various
metal components of the spectrograph  are temperature sensitive and thus are
responsible for the zero-point shifts. Frequent monitoring of spectrograph 
temperature in future should improve our understanding of the zero-point 
shifts.

\clearpage

\begin{figure}[hbtp]
\centering
\includegraphics[width=7cm,angle=270]{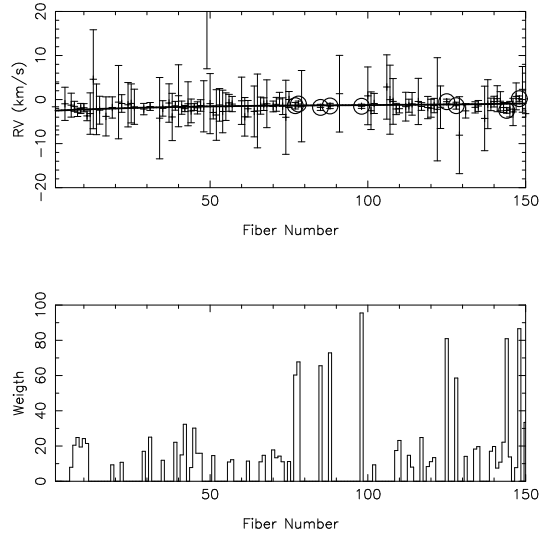}
\caption{Illustration of the procedure  for zero-point calibration.  The top
panel shows  the measured radial velocity for  the sky lines in  each of the
spectra in  one RAVE  field, along with  their errors. Circles  indicate the
location of usable dedicated sky fibers  (with a $R$ value above a threshold
of 5). Sky  fibers with an cross-correlation $R$ value  below the limit are
discarded  from  the   fit  (weight  set  to  0).   Bottom  panel:  weights,
proportional to the value of the  Tonry-Davis $R$ value, used by the fitting
procedure.  Usable dedicated sky fibers have their assigned weight doubled.}
\label{f:zeropoint}
\end{figure}

\clearpage
Given that we  do not know the temperature change between  a given arc frame
and  the  observation frame  to  be  calibrated with  the  arc,  we need  to
establish the radial velocity zero-point  independent of the arc frames.  We
therefore  developed a  procedure  using  the sky  emission  lines that  are
visible  in   the  RAVE  wavelength   region.   We  carry  out   a  standard
cross-correlation technique with the non sky-subtracted target frames, again
using the median spectrum for each fiber from the set of $\geq 5$ scientific
exposures.    We   use   a    sky   spectrum   from   the   Osterbrock   Sky
Spectra,\footnote{www-mpl.sri.com/NVAO/download/Osterbrock.html.}    degraded
to RAVE's  resolution, as a template and  compute, for each  fiber, the radial
velocity associated  with the  sky in that  fiber.  As  the SNR for  the sky
lines is  low, this measurement is precise  only to a few  km/s (the average
radial  velocity  error for  sky  spectra  is  6~km/s). Therefore  a  direct
correction of  a given star's radial  velocity using the  measurement of the
sky zero-point  in the sky+star spectrum  cannot be made. We  instead use the
fact that across  a frame, the expected zero-point  variation is smooth from
fiber to  fiber, so that a  better zero-point estimation can  be obtained by
fitting a low-order polynomial to the \{RV,Fiber\} data set.  Also, in order
to give  more weight in the  fitting procedure to the  dedicated sky fibers,
and to  the fibers containing  better sky spectra  (in the sense  that their
correlation  with  the template  leads  to  more  accurate velocities),  the
individual weight follows  the Tonry-Davis $R$ value.  Further,  to reject all
unreliable measurements from the  fit, fibers with a correlation coefficient
lower than 5 are discarded. Finally, the weight for the dedicated sky fibers
is doubled,  to give an  even stronger constraint  using the fibers  with no
stellar spectra.  Following  this procedure enables us to  estimate the zero
point to $\sim$1~km/s.  An example of this calibration procedure is given in
Fig.~\ref{f:zeropoint}.  The top panel  shows the measured radial velocities
of the sky lines and their  associated errors. The thick line is the adopted
polynomial fit and circles indicate the location of dedicated sky fibers. It
is  clearly seen  that  pure  sky fibers  have  a much  lower  error on  the
determination of the radial velocity, convincing us that more weight must be
given to those fibers in the fitting procedure.  The bottom panel represents
the array  of weights  used to calculate  the polynomial correction  for the
zero point.

\begin{figure}[hbtp]
\centering
\includegraphics[width=7cm,angle=0]{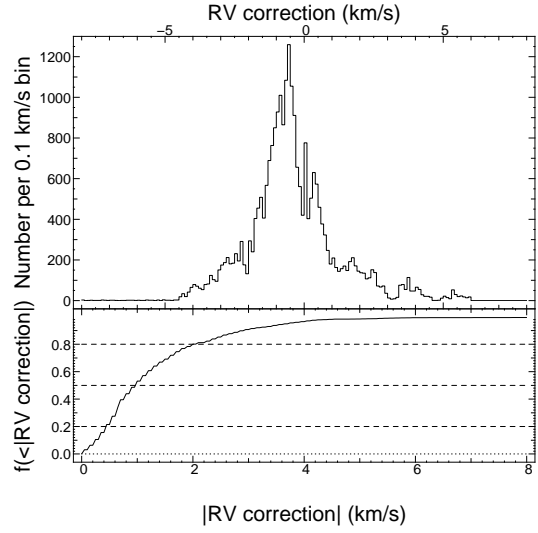}
\caption{Top: histogram of the distribution of radial velocity correction 
due to zero-point offset. The peak value is at $-0.55$~km/s. Bottom: 
fraction of RAVE spectra with an absolute value of this correction lower than 
a given value. Dotted lines show limits at 20\%, 50\%\ and 80\%. 
}
\label{f:rvcor}
\end{figure}

Fig.~\ref{f:rvcor} summarizes the statistics of the applied zero point 
corrections. The upper panel presents their histogram with 0.1~km/s 
bins. The lower panel gives fraction of RAVE spectra with absolute 
correction lower than a given value. The absolute value of the applied 
zero point correction is lower than 2~km/s for 80\%\ of the data.  
\\

\clearpage

\clearpage

\clearpage

\section{Data quality}
\label{s:data_quality}

\subsection{Signal to noise and spectral resolution}
\label{s:signal_to_noise}

\begin{figure}[hbtp]
  \centering  
  \includegraphics[width=7cm]{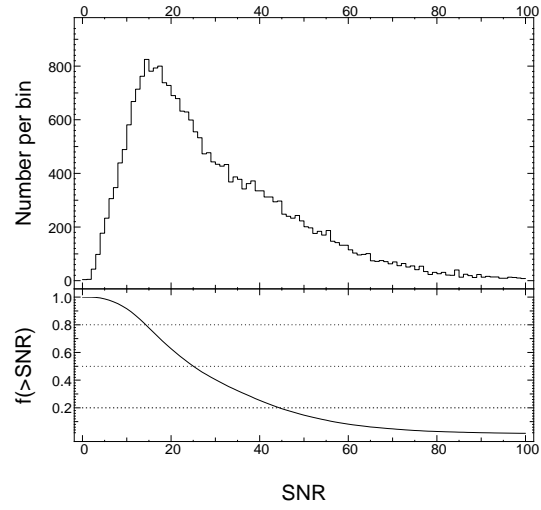}  
  \caption{Top : histogram of the  distribution of signal to noise per pixel
  of  the final one-dimensional  spectrum.  The  peak value  is SNR$\sim$15
  while the mean  value is 29.5. Bottom : fraction of  RAVE spectra with SNR
  larger than  a given value.   Dotted lines show  limits at 20\%,  50\% and
  80\%.  } \label{f:SNR_histogram}
\end{figure}

The signal  to noise ratio for  a given stellar spectrum  was estimated from
the scatter  of the individual exposures  in the sequence  of five 10-minute
exposures of  a given  field. At each  wavelength the scatter  of individual
spectra scaled by their mode  was determined, omitting the 2 most discordant
points.   This   scatter  then  relates   to  the  expected  error   of  the
median-combined  spectrum.   The  reported   SNR  per  pixel  in  the  final
extracted, one-dimensional  spectrum is then  calculated as an  average over
all wavelengths for a given star.  Typical counts per pixel of the final 1-D
spectrum, and per hour of exposure time, for RAVE DR1, follow the
approximate relation
$$N_\R{counts}=10^{-0.4 (I_\R{DENIS}-20.5)}\mbox{,}$$  giving for a complete
exposure  $\sim$33k counts for  an $I$=9  star and  $\sim$2k counts  for an
$I$=12   star.     Note   that   these   SNR    measurements,   reported   in
Fig.~\ref{f:SNR_histogram}, are  conservative: the SNR  in the middle  of the
spectral  domain is  generally better  than the  average, which  is  what is
quoted here. Also all calculations were  done on spectra that were {\it not}
normalized.  Thus  a varying spectral  slope and/or spectral  defocusing can
also worsen the reported SNR. We  estimate that the accuracy of the reported
SNR is $\sim$10\%.

The  resolving  power  $R =  \lambda  /  \Delta  \lambda$ is  obtained  from
measurements of the  width of the emission lines in  the arc exposures. Near
the center of the wavelength range these widths are around 3.1 pixels, which
at $\lambda = 8,600$~\AA \mbox{}  translates into an average resolving power
$R$$\sim$$7,500$. For  stars observed  with the first  $\sim 30$  fibers the
resolving power  is $\sim25$\% lower at  the edges of  the wavelength range,
due to spectral defocusing.  These  values were confirmed by fitting a large
set of  observed spectra  of cool  stars with a  set of  synthetic templates
degraded to different spectral resolutions.

\clearpage

\clearpage

\subsection{Radial velocity accuracy}
\label{s:radial_velocity_accuracy}

As  noted above, the  spectra corresponding  to the  first data  release are
contaminated by second order light, due to the lack of an OG531 filter.  The
main effect  of this contamination is that  it makes it harder  to model the
spectra, as  the amount of contamination  is not known {\it  a priori\/} and
varies  both with  the  spectral type  and with  time,  as the  blue to  red
transmission of a given fiber  changes.  The radial velocities, fortunately, 
are
largely  immune  to this  contamination  by  second-order  light, since  the
location of  the strong absorption lines  in the first-order  spectra -- the
primary determinations  of the wavelength shift in  the cross-correlation --
are not  affected.  The contamination does  lead to a  poorer template match
than would be  achieved for pure 1st-order spectra, but  again in most cases
this results only in a few tenths  of a km/s increase in the internal errors
(as  judged from repeat  observations of  the same  star after  the blocking
filter was inserted, and hence without the 2nd order contamination).

\begin{figure}[hbtp]
  \centering
  \includegraphics[width=7cm]{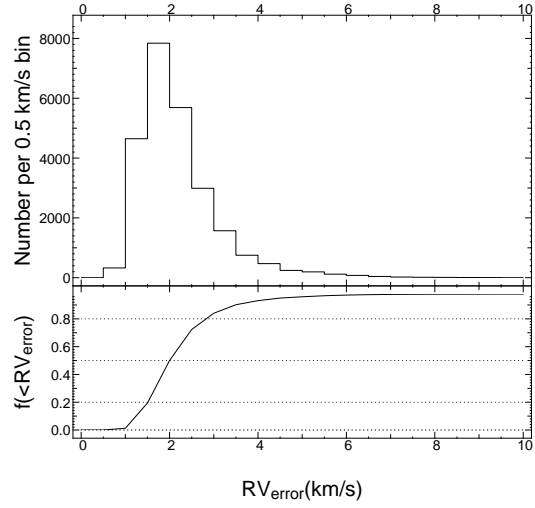}
  \caption{Top panel: distribution of the internal radial velocity error for
    RAVE DR1.  The peak value is at 1.7~km/s while the average error for the
    catalog  is 2.3~km/s.  Bottom panel:  fraction  of RAVE  targets with  a
    radial  velocity error  lower  than  a given  value.   The dotted  lines
    indicate limits of 20\%, 50\% and 80\%.}
  \label{f:eRV_hist}
\end{figure}

The  distribution of  the internal  radial velocity  errors is  presented in
Fig.~\ref{f:eRV_hist}.   The top  panel shows  the histogram  of  the radial
velocity error  in 0.5~km/s bins, while  the bottom panel  is the cumulative
distribution.   The mean internal  error is  2.3~km/s, with  a peak  value of
$\sim$1.7~km/s.   The  bottom  panel  shows  that more  than  80\%  of  RAVE
measurements have an  internal accuracy better than 3~km/s,  and half of the
data released reach an accuracy better than 2~km/s.

\begin{figure*}[hbtp]
  \centering                             
  \includegraphics[width=7cm]{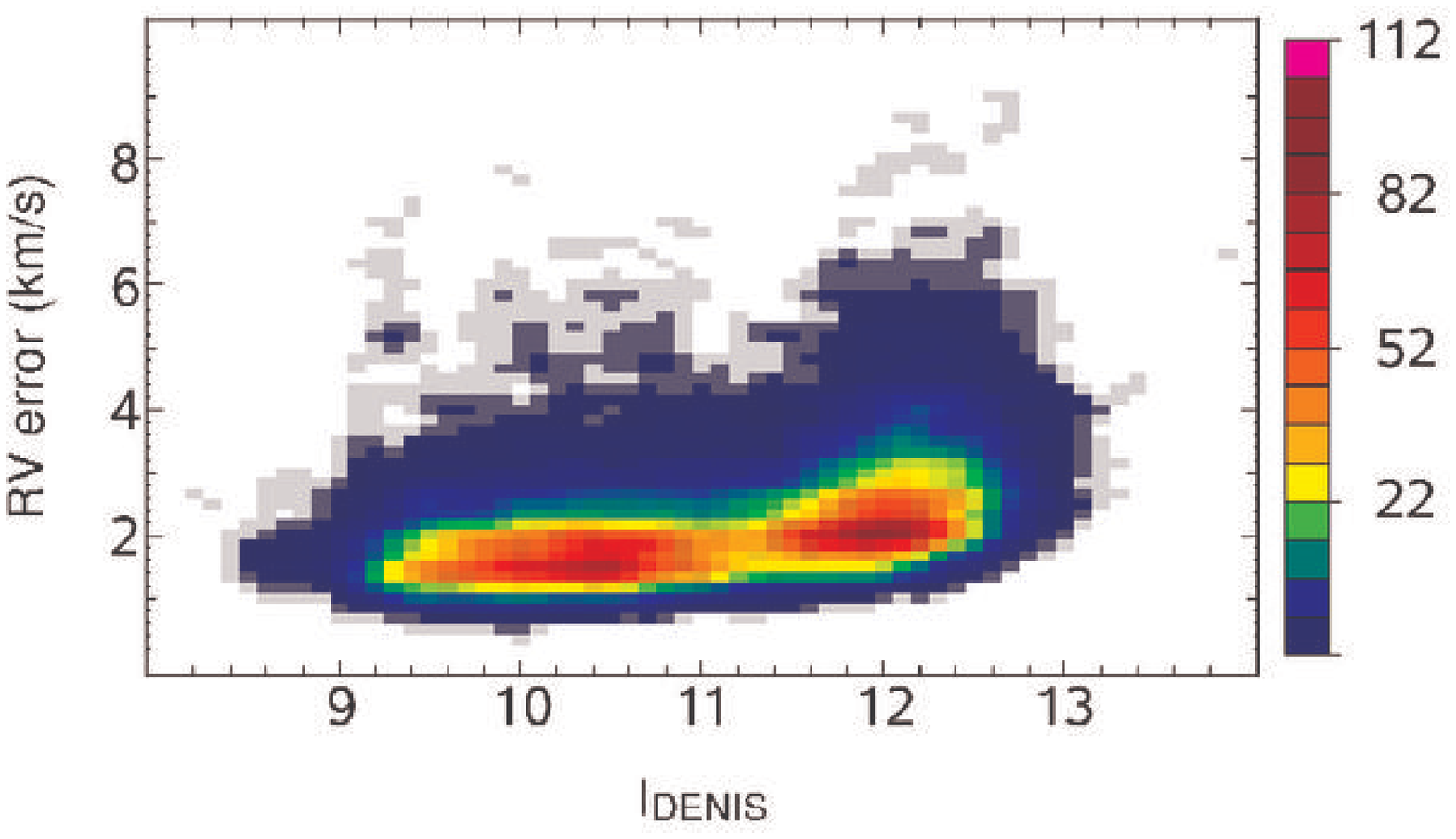}
  \includegraphics[width=7cm]{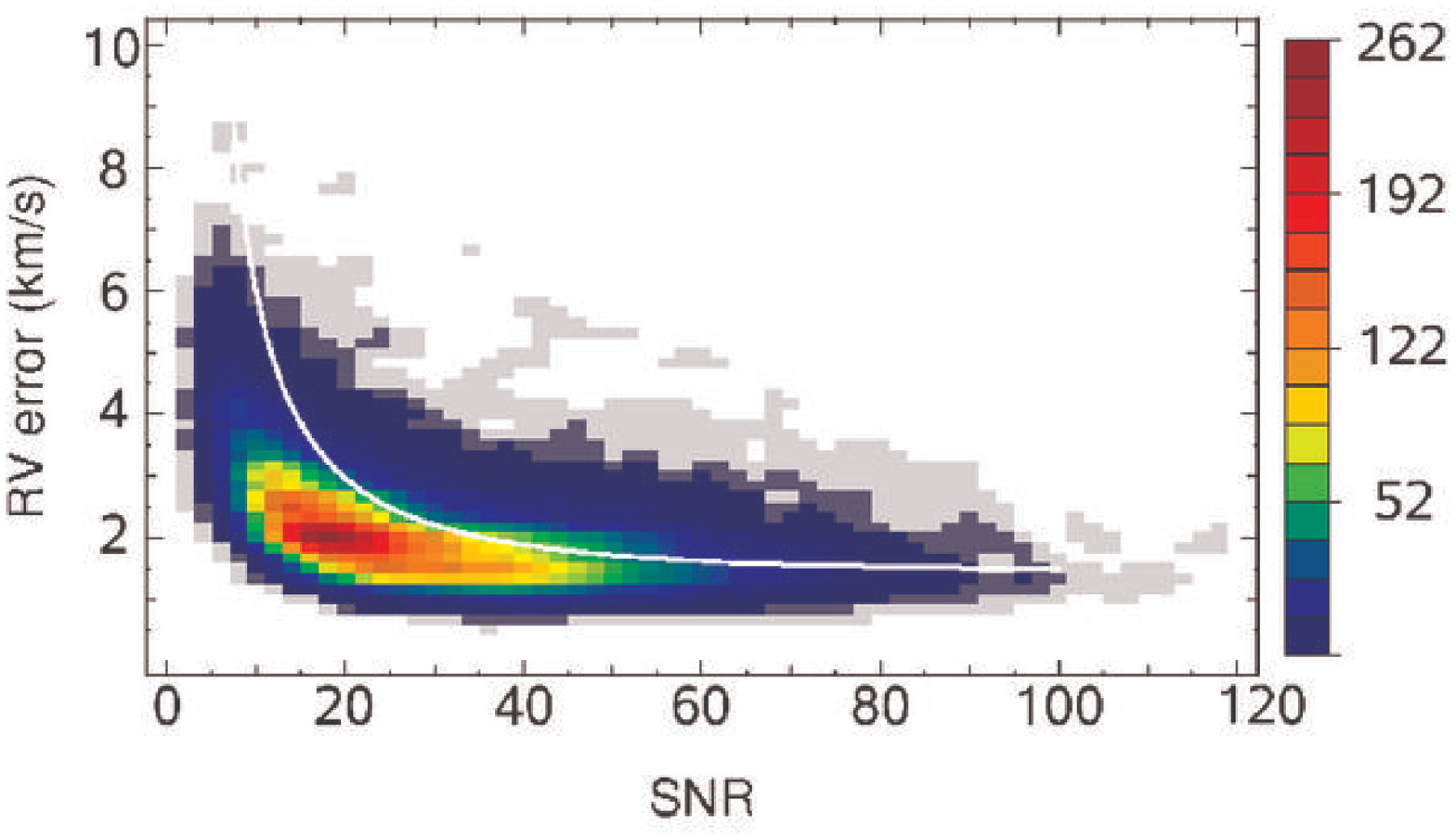}   
  \includegraphics[width=7cm]{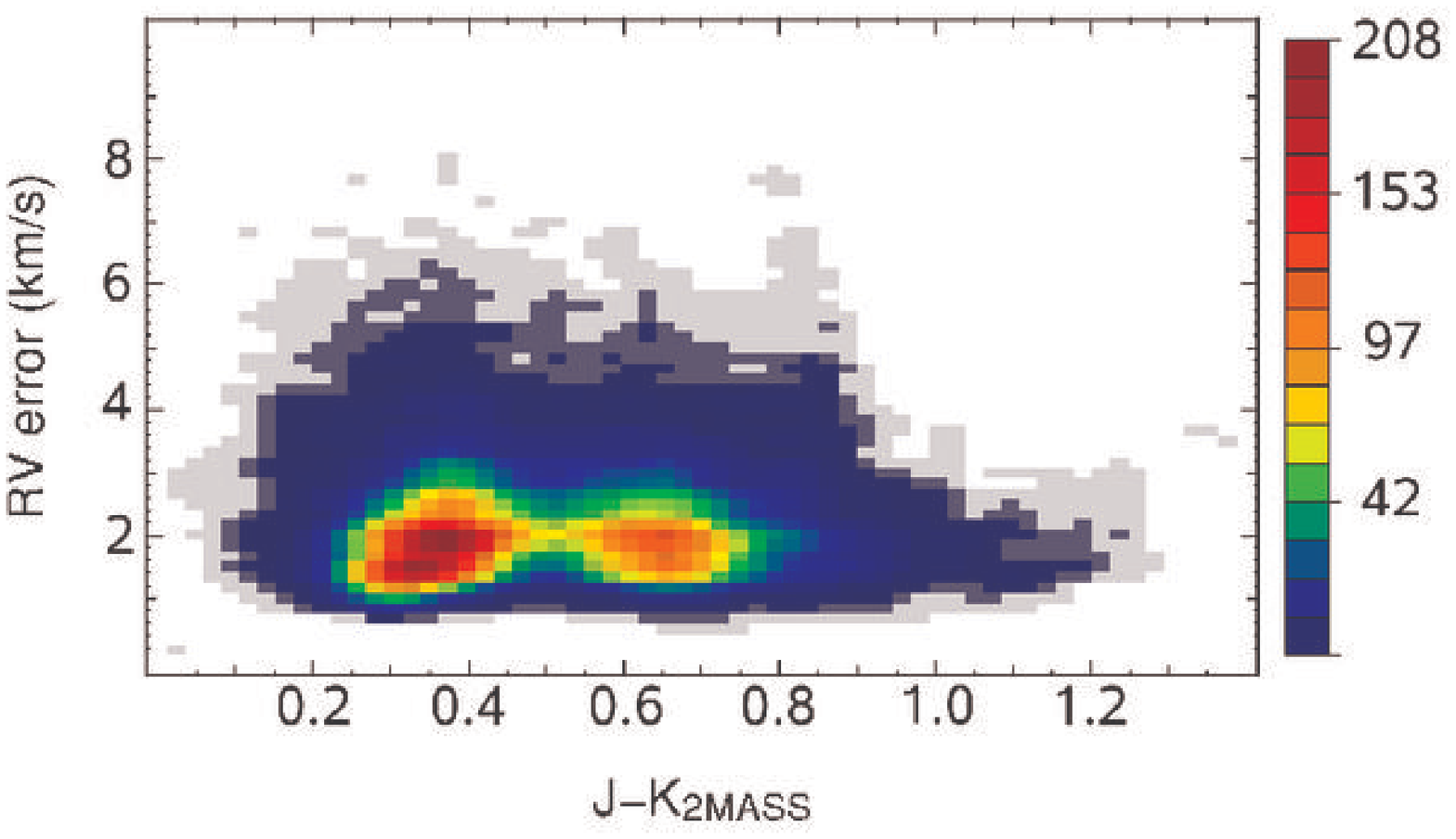}
  \includegraphics[width=7cm]{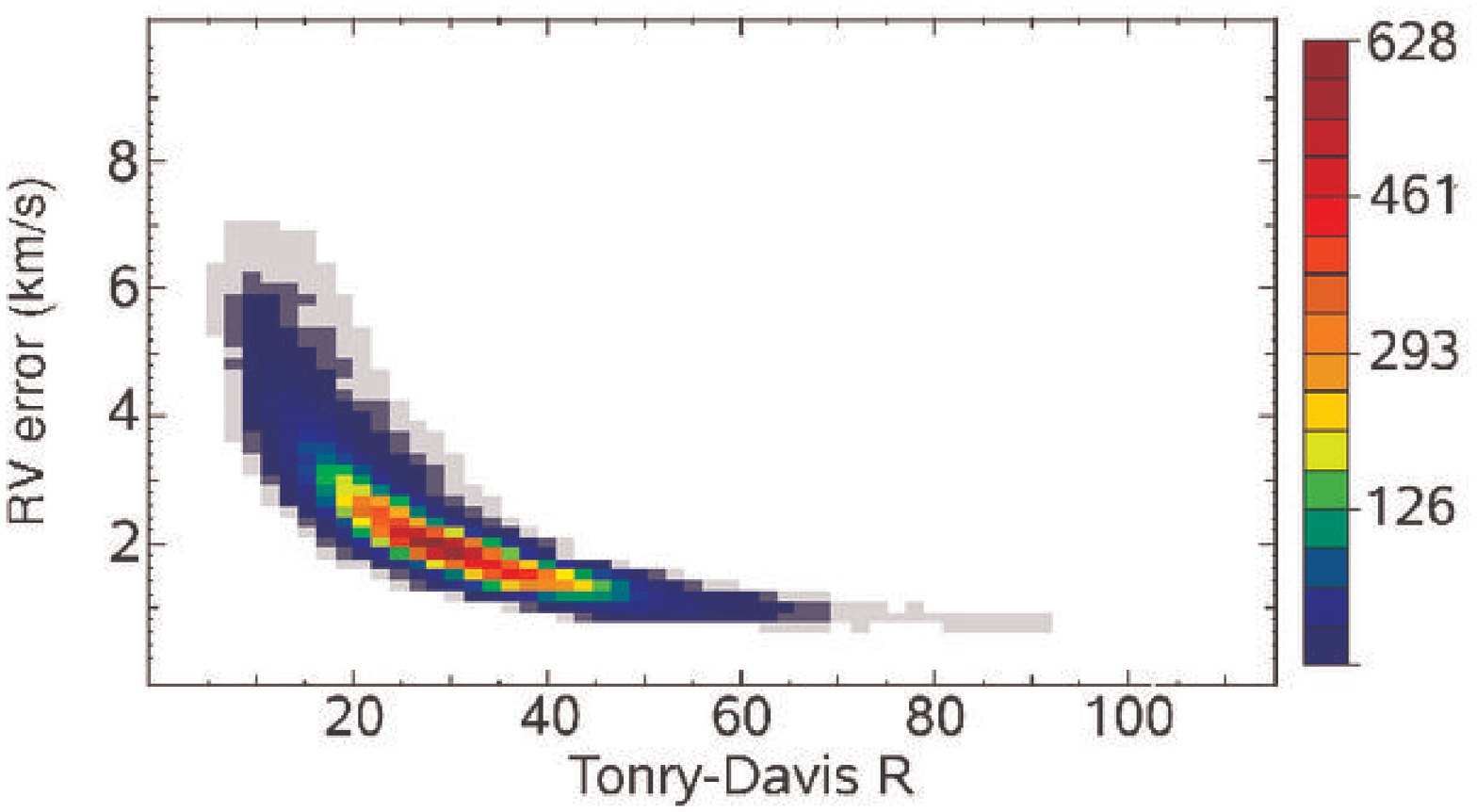}   
  \caption{2D histograms of the distribution of the radial velocity internal
  error. The color  coding follows the number of objects per cell, from blue
  to red.  Top left : as a  function of $I$ magnitude; top right : signal to
  noise ratio --  the white line follows equation  \ref{SNeq}; bottom left :
  $J-K$ color;  bottom right :  Tonry-Davis cross-correlation $R$  value.  }
  \label{f:RVerror}
\end{figure*}

Radial velocity errors achievable for  field stars of F-K spectral type have
been studied during preparations for  the Gaia mission. Munari et al.~(2001)
made observations of IAU radial velocity standards and derived the following
relation for external radial velocity error in km/s as a function of SNR and
spectral resolving power $R$
\begin{eqnarray}
\log (RV\ error) &=& 0.6 (\log SNR)^2 - 2.4 \log SNR \nonumber \\
 & & - 1.75 \log R + 9.36 
\label{SNeq}
\end{eqnarray}
Its  accuracy has  been  confirmed also  by  extensive simulations  (Zwitter
2002).  The error predicted for  $R=7,500$ and SNR$= 30$ (typical values for
RAVE spectra) is 2.2~km/s. Relation~\ref{SNeq} is marked by a white curve of
the  top-right panel in  Fig.~\ref{f:RVerror}. The  fact that  RAVE performs
marginally  better than  predicted can  be  traced to  several factors:  its
wavelength range is a bit wider than for Gaia, its sampling is higher than 2
pixels  per   resolution  element,  blue   order  contamination  contributes
additional spectral  lines carrying velocity information, and  our SNR ratio
estimate is quite conservative.

\clearpage

\clearpage

\begin{figure}[hbtp]
  \centering
  \includegraphics[width=7cm]{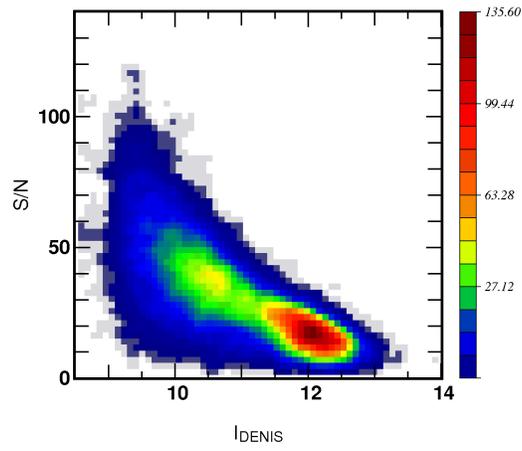}
  \caption{Signal to noise ratio vs DENIS I for RAVE spectra. Again the color coding follows the number of object per bin.}
  \label{f:snr}
\end{figure}

Fig.~\ref{f:RVerror}  presents  the  relation  between the  radial  velocity
internal  error and various  parameters that  can influence  this accuracy.
The upper  left panel  presents the 2D  distribution of the  radial velocity
error  as a  function of  the DENIS  $I$-band magnitude.   The  color coding
follows the  number of objects per cell  from blue (low) to  red (high).  It
can be seen  that the accuracy of the  velocity determination decreases for
stars  with  fainter  $I$--band  magnitudes.  Nevertheless,  there  is  good
consistency between  bright and  faint targets, with  the peak value  of the
error varying  only from $\sim1.4$~km/s  to $\sim2$~km/s between $I= 9$ and
$I=12.5$.   The  increase in  scatter  to  fainter  magnitudes reflects  the
generally   lower  signal-to-noise   at   these  magnitudes,   as  seen   in
Fig.~\ref{f:snr}; the average signal-to-noise  is $\sim 10$ for the faintest
stars and  $\sim 70$ for  the brightest ones.   It should be noted  that the
velocities for  the vast  majority of  the stars, even  at the  faint limit,
remain very accurate.

\clearpage

\clearpage

The effect of Poisson noise is more clearly seen in the upper right panel of
Fig.~\ref{f:RVerror} where  the internal radial velocity  error is presented
as a function of the signal to noise.  This shows that the errors are stable
above a  threshold value in  SNR of around  20, while below this  value, the
velocity errors increase.   To quantify the effect, the  median value of the
internal error, per bin of size 10  in SNR, varies by only 0.3~km/s (peak to
peak) above SNR$=$20,  while the lowest two bins  respectively are larger by
1.1~km/s and 0.5~km/s.  This is also  seen in the bottom right panel, as the
radial velocity error clearly drops as the cross-correlation $R$ coefficient
decreases.  Finally, the bottom left  panel presents the distribution of the
radial velocity error as a function  of the $J-K$ color.  No clear variation
for the peak value is seen in this diagram, but again there is some increase
in scatter  from red  to blue.   This is due  to the  emergence of  the wide
Paschen lines  in the spectra  of early-type stars, and  the correspondingly
wider correlation function peak.

\subsection{Accuracy of the zero point in radial velocity}
\label{s:zeropoint}

The  discussion  above  concerned   the  internal  accuracy  of  the  radial
velocities.  As mentioned in Section~\ref{s:zero_point}, the radial velocity
pipeline  also  corrects the  measurements  for  zero  point offsets.   This
correction is computed  for each spectrum using a  fit to velocities derived
from the sky emission lines.  This process is of course not exact and adds a
further error  term which  must be  taken into account  when using  the RAVE
catalog. The  combined error  (zero-point error $+$  internal error)  on the
radial velocity is discussed later  when comparing RAVE radial velocities to
external data (see section~\ref{s:external_datasets}).

The accuracy of  the zero-point correction is summarized  in the DR1 catalog
in  the {\sc  ZeroPointQualityFLAG} (see  Appendix~A).  This  flag  is built
using a succession  of three characters.  The first  character describes the
dispersion  between  the measured  sky  radial  velocities  and the  adopted
correction (using  the fibers that were  used for the fit)  for the complete
field (therefore a  field flag).  This character ranges from A  to E and the
corresponding dispersion intervals  are reported in Table~\ref{t:zeropoint}.
As  the quality of  the fit  for an  individual spectrum  may depend  on the
location on the CCD, the field  is divided in three equal parts according to
the fiber number  (fiber 1$-$50, 51$-$100 and 101$-$150)  and the dispersion
is   computed    for   each   sub-group    of   fibers   (group    flag   in
Table~\ref{t:zeropoint_stat}).  The second  character in the flag summarizes
the dispersion for the group to  which the fiber belongs (using a flag value
from A to  D in Table~\ref{t:zeropoint}).  Finally, if  the interval between
two successive fibers  with usable sky velocities is  larger than 15 fibers,
the  zero point correction  may not  be well  constrained, even  without the
value of the  dispersion being high.  For those  targets, the last character
of  the flag is  set to  `*', indicating  a possible  zero-point calibration
problem.

\begin{deluxetable}{crll}
\tablecaption{Description of the zero-point quality flag.\label{t:zeropoint}}
\tablewidth{0pt}
\tablecolumns{4}
\tablehead{
\colhead{Flag Value} & \multicolumn{3}{c}{Dispersion}\\
}
\startdata
A & 0~km/s & $< \sigma <$    & 1~km/s\\
B & 1~km/s & $\leq \sigma <$ & 2~km/s\\
C & 2~km/s & $\leq \sigma <$ & 3~km/s\\
D & 3~km/s & $\leq \sigma$    &\\
E & \multicolumn{3}{c}{Less than 15 fibers}\\
& \multicolumn{3}{c}{available for the fit}\\
\enddata
\end{deluxetable}

Table~\ref{t:zeropoint_stat} summarizes the fraction of targets with a given
flag value for each part of the zero-point correction. Over 97\% of the RAVE
targets  have a  zero-point  calibration accurate  to  better than  2~km/s,
$\sim$73\% to  better than 1~km/s. From  those numbers we  can conclude that
our procedure to correct from zero-point offset is efficient, enabling us to
keep the  zero-point error  term below 2~km/s  for the vast majority  of our
targets.

\begin{deluxetable}{ccccc}
\tablecaption{This  table presents  a  summary  of the  zero  point quality  flag
    frequency distributions  in  RAVE first  data  release  for  each flag  group  (see
    text).\label{t:zeropoint_stat} }
\tablewidth{0pt}
\tablehead{
\multicolumn{5}{c}{Field Flag}\\
}
\startdata
A & B & C & D & E\\
\hline
73.7\% & 23.7\% & 1.1\% & 0.4\% & 1.1\%\\
\hline
& & & &\\
\hline
\hline
\multicolumn{5}{c}{Group Flag}\\
\hline
A & B & C & D & \\
\hline
73.8\% & 23.8\% & 1.8\% & 0.6\% &\\
\hline
& & & &\\
\hline
\hline
\multicolumn{5}{c}{Possible Bad Calibration}\\
\hline
&  & * &  & \\
\hline
&  & 3.7\% &  &\\
\enddata
\end{deluxetable}

Adopting a zero-point error of 1~km/s, the mode of the radial velocity error
distribution for the RAVE DR1 catalog is 2~km/s.

\subsection{Repeated observations}

A total  of 428 stars  in the present  data release were observed  more than
once during  the first year of  operation. Of these, 98  stars were observed
three times.

\begin{figure}[hbtp]
  \centering
  \includegraphics[width=7cm]{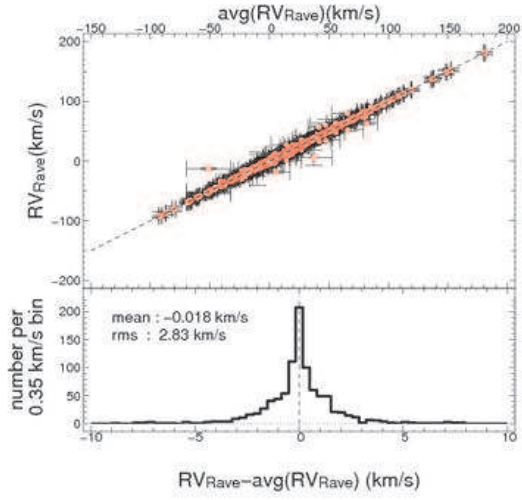}
  \caption{Radial velocity accuracy checked  by repeated observations of 428
    stars.   The  scatter of  2.83~km/s  is  consistent with  estimates  of
    internal and zero-point RV errors.}
  \label{f:reobs}
\end{figure}

Assuming  that  any variation  of  the  measured  radial velocities  is  not
intrinsic,  one  can  thus  estimate  the  accuracy  of  the  catalog.   The
distribution of the derived radial  velocities of the repeat observations is
shown in  Fig.~\ref{f:reobs}.  The mean  deviation is essentially  zero, and
the rms  of 2.83~km/s  is perfectly consistent  with the internal  error and
zero-point accuracy given above.

\clearpage

\clearpage

\subsection{Contamination by binary stars}

Most of the stars in the present  data release are observed only once, so we
have  no way  of knowing  whether a  given star  is single,  or  whether the
observed  radial  velocity may  in  part  reflect  the instantaneous  radial
component of  the orbital  motion in a  binary system.  The  average orbital
velocity  of star 1  in orbit  around star  2 is  $$<V_\R{orb}> =  v_o\, q\,
(1+q)^{-1}\, (M/P)^{1/3}\mbox{,}$$ where $v_o = 30$~km~s$^{-1}$, $q=M_2/M_1$
is  the mass ratio,  and $M$,  $P$ are  the total  mass and  orbital period,
expressed in solar masses and years for the resultant velocity in km/s. Thus
for most single-lined systems, with an  orbital period of years or less, the
orbital radial  velocity projection exceeds 5~km/s and  their kinematic data
may be inaccurate and even  potentially misleading.  This is consistent with
the 15 km/s value for the average scatter of individual radial velocities of
spectroscopic binaries found by  \citet{copenhagen}. Note that these authors
made  many repeat  observations  and  identified 19\%\  of  their sample  as
spectroscopic binaries of some  kind.  Double-lined systems show asymmetric,
or  even double-peaked,  shapes of  the correlation  function.   The present
release contains  27 targets  flagged as possible  binaries since  they show
clear  asymmetric profiles  for the  correlation  function as  well as  poor
template  match,  or clear  double-line  spectra.   Nevertheless, we  cannot
unequivocably state  these are binaries without  temporal orbital variations
to the  radial velocities, as  non-intrinsic effects (for  example scattered
light residual  or fiber cross-talk) could also  contribute.  Therefore this
information should be used with caution.

Photometric variability is an efficient  way to discover binary systems with
periods  of hours  or days  \citep{picard},  while systems  with periods  of
hundreds of years are spatially resolved  (see Figs.~3 and 4 from Zwitter \&
Munari  2004). Unfortunately  no repeated  photometric observation  or light
curves are available for stars in this data release.  A cross-check with the
general catalog  of variable stars \citep{gcvs} revealed  22 matches.  Those
objects  are flagged  in  the catalog  using  the {\sc  VarFLAG} entry  (see
Appendix~A).  While every  effort has been made to  identify visual binaries
from  available catalogs,  as well  as at  the telescope,  we note  that the
amplitudes of the orbital velocities of member stars of such systems is very
small, and indeed  less than our internal accuracy, so  would not affect our
quoted radial velocity values.

In future  years repeats will  be obtained for  a larger fraction  of stars,
including  some of  the stars  from this  data release.   This will  help to
identify spectroscopic binaries and active stars.

\subsection{Comparison with external datasets}
\label{s:external_datasets}

To further  check the consistency of  our radial velocities,  we compare our
measurements with external data from three main sources\footnote{A few stars
from this  release were also observed  with the echelle  spectrograph at the
APO 3.5m telescope, as part of a follow-up scientific project.  They will be
published as part  of the science paper.}: dedicated  observations using the
ELODIE spectrograph; dedicated observations  using the MSSSO 2.3m telescope;
overlap with the sample of the Geneva-Copenhagen survey (GCS).  Each sub-set
will be briefly described below.

\paragraph{ELODIE data:}\mbox{}

ELODIE is a cross-dispersed \'echelle spectrograph mounted at the Cassegrain
focus of  the 1.93m  telescope at the  Observatoire de Haute  Provence.  The
spectra cover a $3,000$~\AA \mbox{} wavelength range ($3,850\!-\!6,800$~\AA)
with a spectral  resolution of 42,000.  The instrument  is entirely computer
controlled  and   a  standard  reduction  software   package  ({\sc  TACOS})
automatically  processes  the data.   Seven  stars  of  this first  release,
selected simply to  be bright and accessible, were  observed on three nights
in July 2004 (July  5, 8 and 9).  The SNRs of the spectra  range from 1 to 5
per pixel  (10 to  20 minutes  of exposure time),  sufficient at  these high
spectral  resolutions\footnote{An  accuracy of  10~m/s  can  be reached  for
planet searches  using this  instrument.}.  We cross-correlated  the spectra
against the most appropriate spectral template given by the ELODIE reduction
pipeline\footnote{TACOS provides two templates, the first one {\sc R37000K0}
is valid  for G, K  and M  stars while {\sc  R37000F0} is used  for spectral
types    around   F0.}.     The    results   are    given   in    Appendix~B
Table~\ref{t:external_ELODIE}.  The limiting accuracy ranges from 300~m/s to
1~km/s,    with    the   exception    of    the    probable   double    star
(\mbox{TYC~5031\_478\_1}) that  has been discarded from  the analysis.  From
the six  single stars measured with  ELODIE for this release,  we obtained a
dispersion  $\sigma(RV_{\rm ELODIE}-RV_{\rm  RAVE})=1.7$~km/s,  dominated by
errors in the RAVE measurements.  We  find a negligible offset with the mean
being $<(RV_{\rm ELODIE}-RV_{\rm RAVE})> =-0.1\pm0.8$~km/s, showing that the
final zero-point  error for RAVE data  is much smaller  than the measurement
dispersion.  \\

\paragraph{2.3m data:}\mbox{}

Long  slit  spectra were  taken  with  the  Double Beam  Spectrograph  (DBS)
\citep{rodgers} on the MSSSO 2.3m telescope, over a period of two years from
December 2003 to September 2005.   The average seeing was $1.5''$ - $2.5''$,
where  a  narrow  slit  ($1.5''$)  minimized slit  positioning  errors.  The
spectral   resolution    is   similar   to    RAVE   (R$\sim$$8,000$)   with
$0.55\textrm{\AA}/\mathrm{pix}$.   The  stars,  randomly selected  from  the
first  year  of  observation  with  the exception  of  a  few  high-velocity
candidates, were observed at a vertical angle of $0^\circ$ or $180^\circ$ to
eliminate atmospheric dispersion effects.   The average SNR was $\sim$90 per
pixel.   We  cross-correlated  these   spectra  against  a  sub-set  of  the
\citet{zwitter04} spectra using XCSAO in IRAF.  The best match was chosen as
the  template with the  highest $R$  coefficient \citep{tonry}.   Only those
stars with a best-match $R>40$ were  used - this limit was found empirically
to minimize errors in 2.3m  radial velocities.  The resulting internal error
of  the  2.3m  radial   velocities  from  these  methods,  measured  against
high-precision Nordstr\"om  et al.~stars  and radial velocity  standards, is
$\sim$1.5~km/s.      The     results     are     given     in     Appendix~B
Table~\ref{t:external_2p3}.

The  77 stars  belonging to  this data  release and  measured with  the 2.3m
telescope cover a radial velocity range  of 820~km/s. Of those, we believe 7
outliers are  likely single-lined spectroscopic binaries,  as the difference
in radial velocity  between the RAVE and 2.3m  measurements deviates by over
$3\sigma$.       Excluding      these      7     outliers,      we      find
$\sigma(RV_\textrm{2.3m}-RV_\R{RAVE})=3.4$~km/s                           and
$\mu(RV_\textrm{2.3m}-RV_\R{RAVE})=0.1\pm0.4$~km/s.    Correcting  for  2.3m
internal error  we conclude from this  sample that the  RAVE radial velocity
error is $\sigma=3.0$ km/s, with a negligible mean offset.

\paragraph{Geneva-Copenhagen targets:}\mbox{}

DR1 targets have been cross-matched with the GCS catalog, which by virtue of
many  repeated  observations contains  binarity  indicators  in addition  to
accurate radial velocities.  We found  13 matches, two of them classified as
binaries in the GCS.

The  resulting 11  `single' targets  show acceptable  agreement with  a mean
difference of $1.4\pm0.4$~km/s and a  rms of 1.4~km/s. The results are given
in Appendix~B Table~\ref{t:external_Copenhagen}.

Figs.~\ref{f:external_CCD}  to~\ref{f:external_JK} summarize  the comparison
of  RAVE RVs with  the external  data.  Fig.~\ref{f:external_CCD}  shows the
location in the 2MASS color-color  diagram of the RAVE targets with external
measurements.   The broad  distribution  of those  objects  in this  diagram
indicates that we have good  coverage of the global stellar properties, with
the  exception  only  of  the  reddest  giants  ($H\!-\!K\gtrsim0.2$).   Our
external checks should  therefore be applicable to the  full survey, despite
the low number of external targets.

\clearpage

\begin{figure}[hbtp]
  \centering \includegraphics[width=7cm]{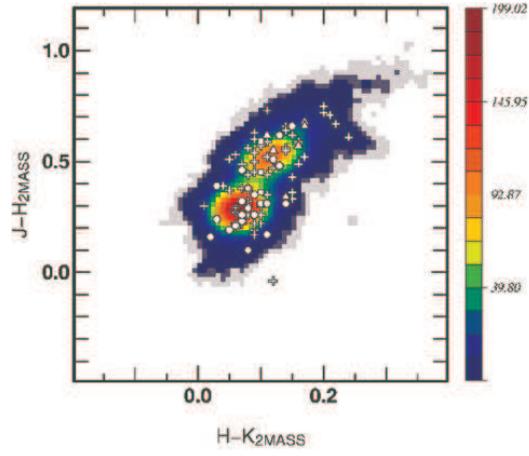} \caption{Infrared
  color-color diagram of the stars in DR1. The locations of our
  targets with external data are overplotted using white symbols. 
  Crosses are for 2.3~m data, triangles for ELODIE targets and squares
  for GCS targets.  Open circles denote the location of binary
  targets.  The external targets are well-distributed across this
  parameter space, indicating that our reference target selection is
  not biased towards any particular type of objects.}
  \label{f:external_CCD}
\end{figure}

\clearpage

Fig.~\ref{f:external_rvcomp} presents the comparison between the RAVE radial
velocities and the  external measurements. The top panel shows  the RV to RV
comparison,  each  different source  plotted  with  different symbols  (open
circles are  for binary  systems that are  removed from the  analysis).  For
each data source, the mean  difference and standard deviation is also given.
As the spread in radial velocity is large, the bottom panel gives the radial
velocity difference as  a function of RV.  This plot  clearly shows that the
binary  systems  produce  a  large  offset, while  single  stars  show  good
agreement,  with  close  to  zero  velocity  difference.   We  concluded  in
Section~\ref{s:zeropoint}  that the mode  is 2~km/s,  the median  RAVE error
being $\sim$2.2~km/s.   Combining the  measurements from all  three samples,
the mean difference between RAVE radial velocities and external measurements
is  $-0.04\pm0.29$~km/s  with a  standard  deviation of  3~km/s\footnote{The
standard   deviation   corrected  for   external   source   mean  error   is
$\sigma=2.7$~km/s.}.  This is in good agreement with our predicted accuracy.

\clearpage

\begin{figure}[hbtp]
  \centering            
  \includegraphics[width=7cm]{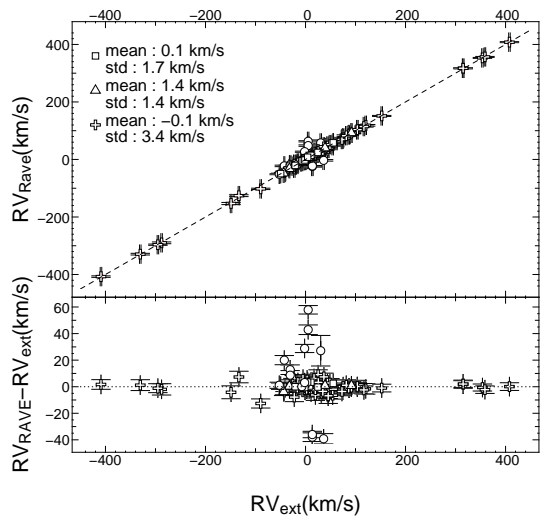}
  \caption{Comparison of RAVE  radial velocities with external measurements.
  The   top  panel   presents   the  external   to   RAVE  radial   velocity
  comparison. The  bottom panel  shows the radial  velocity difference  as a
  function  of radial  velocity for  the three  different  reference groups.
  Squares are for  ELODIE, triangles for GCS and  crosses for 2.3~m targets.
  Open circles denote  suspected and confirmed binary objects  and are removed
  from the analysis. The mean difference and standard deviation are reported
  for  each group  in the  top panel.   RAVE radial  velocities show  a good
  agreement with independent measurements.  }  \label{f:external_rvcomp}
\end{figure}

\clearpage

As  a  further test,  we  compare  in  Fig.~\ref{f:external_SNR} the  radial
velocity difference to the SNR in RAVE data. The top panel in this figure is
a simple scatter  plot including all available data,  while the bottom panel
presents the mean difference and rms  for intervals of 10 in SNR. The offset
in the interval  50$\leq$SNR$<$60 is due to a  single measurement with large
error and  RV difference  (RV$_{diff}=$7$\pm$5~km/s).  This figure  shows no
dependence of the radial velocity on  SNR, the mean and rms being consistent
with no  trend.  Again, the  amplitude of the  scatter increases as  the SNR
drops  below   $\sim  20$,  as  expected   from  the  top   right  panel  of
Fig.~\ref{f:RVerror}.

Finally Fig.~\ref{f:external_JK} demonstrates that radial velocity errors do
not depend on color. The exception  are the bluest of our targets ($J-K\!<\!
0.1$) which  are dominated  by wide  hydrogen lines with  only a  handful of
weak metallic lines.

\clearpage

\begin{figure}[hbtp]
  \centering
  \includegraphics[width=7cm]{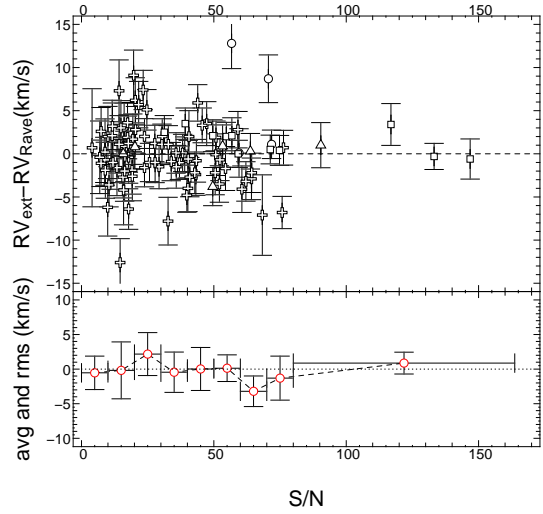}
  \caption{Behaviour of the radial velocity difference with signal to noise.
  Top panel  : radial velocity difference  as a function of  signal to noise
  for   each  group   of   external  measurements.    Symbols   are  as   in
  Fig.~\ref{f:external_CCD}. Bottom  panel : average deviation and  rms as a
  function of signal to noise for the sum of all external measurements. This
  plot shows  no apparent  bias as  a function of  SNR, the  mean difference
  being consistent with zero for  all values of SNR. One bin (60$<$SNR$<$70)
  has a rather large deviation due to one single object with large error.  }
  \label{f:external_SNR}
\end{figure}

\begin{figure}[hbtp]
  \centering
  \includegraphics[width=7cm]{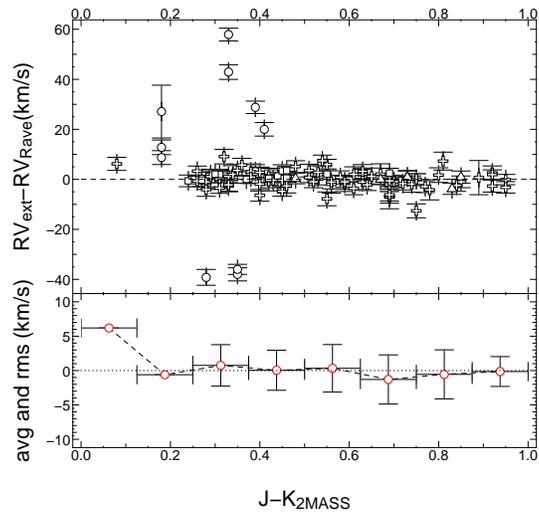}
  \caption{Behaviour of the radial velocity difference with $J-K$ color from 
  the 2MASS catalog. Arrangement and symbols as in Fig.~\ref{f:external_SNR}.
  Radial velocity errors do not depend on color of the 
  target except for the bluest targets. 
}
  \label{f:external_JK}
\end{figure}

\clearpage

In summary,  a comparison of the  RAVE velocities with those  from the three
external sources  (ELODIE, GCS  and 2.3-m) reveals  that the offset  in mean
velocity is  very small, most  likely less than  1~km/s.  The errors  in the
RAVE  stars in  common with  the  ELODIE and  GCS observations  are 1.7  and
1.4~km/s respectively, while the  corresponding error in the RAVE velocities
of 2.3-m stars are larger, about  3.0~km/s.  We note that the ELODIE and GCS
stars  are significantly  brighter in  the mean  than the  2.3-m  stars (see
Fig.~\ref{f:Imag_geneva_rave}).   However,  Fig.~\ref{f:external_SNR}  shows
that the difference in estimated errors is probably not due to different SNR
values  for  the  RAVE spectra.   A  more  likely  explanation is  that  the
contamination from  stars in adjacent  fibers, while small,  will contribute
more to the velocity errors for the fainter 2.3-m stars.

\section{First data release}
\label{s:data_release}

\subsection{Spatial coverage and global properties}
\label{s:coverage_dr1}

The first RAVE data release presents radial velocities for 24,748 apparently
bright stars  in the Milky  Way. The total  number of spectra  collected for
these  stars  is 25,274  (including  re-observations).   Those spectra  were
obtained  during  67 nights  between  April  $11^{\rm{th}}$  2003 and  April
$2^{\rm{nd}}$  2004, with  the exception  of  one field  observed on  August
$3^{\rm{rd}}$   2004   (second    year,   but   affected   by   second-order
contamination).  12,836 stars in this release are Tycho-2 entries and 11,921
are extracted  from the SuperCOSMOS  Sky Survey.  The stars  are distributed
over 235 6dF fields (for 240 observed including re-observations) for a total
area covered of $\sim$4,760 sq.~deg.

\begin{figure*}[hbtp]
  \centering
  \includegraphics[angle=90,width=10cm]{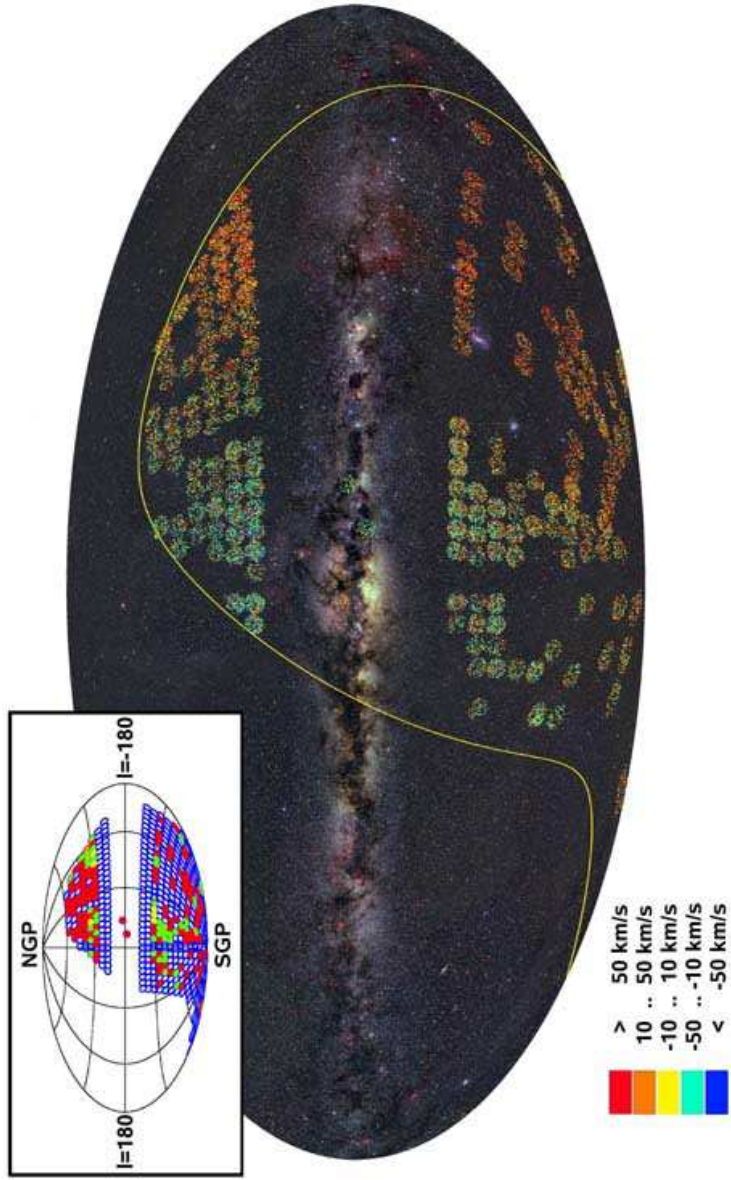}
  \caption{Aitoff  projection in  Galactic  coordinates of  RAVE first  data
    release fields. With the exception of two test fields, all field centers
    are  located  at  $|b|  \ge  25^\mathrm{o}$.   The  heliocentric  radial
    velocity gradient around  the sky, traced by the  colors, reflects Solar
    motion and the  projection of the different components  of velocity down
    the line-of-sight.  The yellow line represents the celestial equator and
    the  background is from  Axel Mellinger's  all-sky panorama.   The inset
    shows the location of the  RAVE fields for the same projection. Observed
    fields are  color-coded according to the  number of times they have been
    visited;  red is for  one pointing,  yellow two,  green three  and light
    brown four.  } \label{f:chart_gal_skypanorama}
\end{figure*}

The  locations on  the  sky  of the  RAVE  fields of  DR1  are indicated  in
Fig.~\ref{f:chart_gal_skypanorama}.    This   figure   presents  an   Aitoff
projection  of all  RAVE DR1  target fields  in Galactic  coordinates.  Each
circle is an RAVE target field (5.7~degrees in diameter).  Filled fields are
part  of  this data  release.   The color  coding  indicates  the number  of
pointings for a  given field (remember that each field  contains up to three
sets of  targets): red  for one time,  yellow for  two and green  for three.
Among those, 4 fields are re-observed  with the same set of targets (3 fields
have been re-observed a second time, one field three times).  In addition to
the normal RAVE  target fields, which are restricted  to $|b|>20^\circ$, two
fields  observed  to test  the  MK-classification  scheme  in the  red  RAVE
wavelength region are located closer to the Galactic plane.

\subsection{Photometry}
\label{s:photometry}

As noted earlier,  the photometry of the input  catalog was not homogeneous:
the  bright part of  the sample  was selected  using a  pseudo-$I$ criterion
based on  Tycho-2 $V_{\rm{T}}$ and $B_{\rm{T}}$ magnitudes,  while the faint
part     was     selected     from     SuperCOSMOS     $I$     data     (see
section~\ref{s:input_catalog}).  To maximize the  usability of the DR1 data,
cross-identification with optical  and near-infrared catalogs (USNO-B, DENIS
and 2MASS) is provided.  The  matches are selected using a nearest-neighbor
association.   A  large  search  radius   is  used  to  check  for  possible
contamination, given that the diameter of  a 6dF fiber on  the sky is
6.7~arcsec.

In  addition  to  the distance  to  the  nearest  neighbor, a  quality  flag
summarizing the reliability  of the association is provided  in the catalog.
This single character  flag is set to \rm{A}  for clear association, meaning
that the distance  between the target and the nearest  neighbor is less than
1~arcsec, with  no other  possible association.  The  flag value  \rm{B} (or
\rm{C}) is used  to warn that two (or more) associations  are found within a
1~arcsec radius,  and \rm{D} indicates  that the nearest neighbor  is further
away  than  2~arcsec.   A  value  \rm{X}  is given  for  the  flag  when  no
association  is  found.  Table~\ref{t:xid}  summarizes  the  outcome of  the
cross-identification procedure for the various photometric catalogs.\\

\begin{deluxetable}{cccccccc}
  \tablecaption{Number and fraction of RAVE targets with a counterpart in the
      photometric catalogs.\label{t:xid}}
\tablewidth{0pt}
\tablehead{
 Catalog Name & Number  of  & \% of DR1 & \multicolumn{4}{c}{\% with quality flag}\\
      & Objects& with Counterpart& A & B & C & D\\ 
}
\startdata  
2MASS &  25,268 & 99.98\% & 99.6\% & 0\% & 0\% & 0.4\%\\ 
DENIS & 18,637& 73.7\% & 75.3\% & 22.3\% & 1.8\%& 0.6\%\\ 
USNO-B & 24,814 & 98.2\% & 99.2\% & 0.5\% & 0\% & 0.3\%\\ 
\enddata
\end{deluxetable}

For 2MASS  and USNO-B, the cross-identifica\-tion appears  well-defined. The
fraction of  objects with a flag  value different from \rm{A}  is lower than
1\%,  indicating good  agreement between  the astrometry  in the  RAVE input
catalog and USNO-B. We therefore expect the level of false identification to
be lower than 1\% for those catalogs.

The result  of the  cross-identification with DENIS  appears worse  than for
2MASS  or USNO-B.  The  main differences  are in  the categories  \rm{B} and
\rm{C}, where  multiple matches are found.   This is due to  problems in the
DENIS  catalog where,  at  the edges  of  the detector,  the astrometry  and
photometry both become less accurate.   In this case, DENIS reports multiple
detections of  the object  for the different  scans.  Those  detections have
almost identical magnitudes  and positions but could not  be associated with
the same object.  In all those  cases, we use the nearest neighbor.  As the
difference in magnitude  is small between the possible  matches, this should
not affect  the overall  quality of DENIS  associations.  Also  the reported
magnitudes should be of sufficient accuracy for most of the uses of the RAVE
catalog.  Therefore, even if the cross-identification with the DENIS catalog
seems to be poorer according to Table~\ref{t:xid}, it does not significantly
lower the validity of the cross-identification. \\

\begin{figure}[hbtp]
  \centering
  \includegraphics[width=7cm]{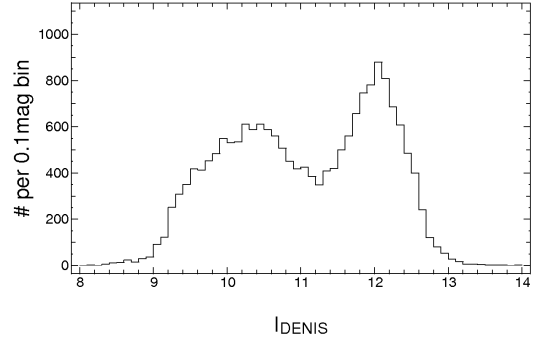}
  \caption{DENIS $I$  band apparent magnitude  distribution of the  RAVE DR1
    catalog.   The two  peaks in  this distribution  are the  result  of our
    selection criteria in  pseudo-$I$ band and delimit the  bright and faint
    samples.}
  \label{f:Imag_hist}
\end{figure}

The  resulting  $I$-band distribution  for  the  RAVE  catalog is  given  in
Fig.~\ref{f:Imag_hist}  using  the  cross-identification with  DENIS.   This
luminosity  function  exhibits two  peaks,  defining  the  bright and  faint
samples: the  first one is centered  on $I=10.2$, the second  one on $I=12$.
The shape of this function clearly  indicates that, at this point, RAVE does
not approximate a random magnitude-limited sample.  As discussed earlier in
Section~\ref{s:input_catalog}, this is an  effect of the input catalog being
selected  either  using  the   pseudo-$I$  magnitude  derived  from  Tycho-2
photometry or using  SuperCOSMOS $I$ in the bright  regime, where systematic
offsets are  to be expected.  This effect will  be corrected in  future RAVE
releases with the availability of the  full DENIS catalog, which will be used
for the input catalog of later data releases.\\

\clearpage

\clearpage

\begin{figure}[hbtp]
  \centering
  \includegraphics[width=7cm]{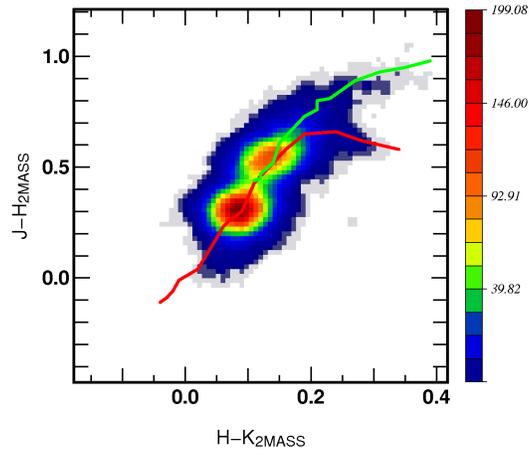}
  \caption{2-MASS infrared color-color diagram for RAVE targets in the first
    data release.   For clarity, the predicted loci for dwarfs and giants
    (resp.  dark and light curves) from \citet{wainscoat} have been added.
Again the color-code indicates the number of objects per bin.}
  \label{f:JH_HK}
\end{figure}

Fig.~\ref{f:JH_HK}  presents  the  2MASS  color-color diagram  of  the  data
release.  Fiducial colors from  Table~2 of \citet{wainscoat} have been added
for  clarity, with  dark and  light gray  curves for  dwarf and  giant stars
respectively.   This clearly  shows  that, as  intended,  the data  released
probes both the nearby and more distant Galaxy.  As an example, K0 dwarfs in
the  RAVE catalog  span  an  approximate distance  range  of $\sim$50~pc  to
$\sim$250~pc, while  K0 giants are located  in the distance  range 700~pc to
3~kpc.\\

\clearpage

\clearpage

\subsection{Proper motions}
\label{s:proper_motions}

The input catalog for the DR1 has  been built upon Tycho-2 and SSS, and both
of  these provide  proper  motions.   When available,  we  also used  proper
motions from  the Starnet~2.0 catalog  which is presently being  compiled at
the Astronomisches Rechen-Institut, Heidelberg.   It is based on the Starnet
catalog of positions and proper motions \citep{Starnet} of 4.3 million stars
based on measurements from the Astrographic Catalog and GSC~1.2
\citep{GSC12}.   The average  rms  error  of proper  motions  in Starnet  is
5~mas/yr, being  mainly influenced by  the accuracy of GSC~1.2.   During the
last  decade new,  highly accurate,  astrometric catalogs  became available,
namely Tycho-2,  UCAC2 and, for  fainter stars, 2MASS.   Starnet~2.0 combines
Starnet with these three  catalogs in a rigorous, weighted least-squares
adjustment to  derive mean  positions and proper  motions.  Weights  for the
individual positions in  the catalogs have been attributed  according to the
accuracy of the  position measurements in the sources:  for the Astrographic
Catalogs   weights   were  taken   from   \citet{TRC},   for  GSC~1.2   from
\citet{GSC12}, for Tycho-2 from  \citet{tycho2}, for UCAC2 from \citet{UCAC}
and for  2MASS from  \citet{2mass}.  The mean  errors of the  proper motions
given in  Starnet~2.0 are calculated  from the individual weights.  They are
not calculated from the dispersion of the residuals due to the low degree of
freedom in each individual adjustment.  The average mean error in the proper
motions  of  Starnet~2.0  stars  in  this  catalog  is  2.6~mas/yr  in  each
coordinate.\\

Neither Starnet nor Tycho-2 or UCAC2  are complete with regard to stars with
proper   motions  in   the   range  between   approximately  30~mas/yr   and
150~mas/yr. This is mainly due to the fact that the epoch difference between
the  Astrographic   Catalog  and   modern  catalogs  is   almost  100~years.
Therefore, the identifications of  these moderately high proper motion stars
is  uncertain.  To overcome  this difficulty  for Starnet~2.0,  an auxiliary
catalog has been constructed from  the difference between 2MASS and GSC~1.2,
with an epoch  difference of about 20~years.  The average  mean error in the
proper motions  of the stars from  this source is  18~mas/yr. Proper motions
are taken from this auxiliary catalog whenever no other proper motions could
be found.  A summary of the contribution of each proper motion source to DR1
is  given in  Table~\ref{t:spm}. The  source of  the proper  motion  is also
flagged in the catalog (see Appendix~A).

\begin{deluxetable}{rccc}
\tablecaption{Summary of the proper motion sources.\label{t:spm}}
\tablewidth{0pt}
\tablehead{
  SPM  & Catalog & Number & \% \\
  Flag & Name & of sources& of DR1\\
}
\startdata
0 & No proper motion & 13 & 0.05\\
1 & Tycho-2 & 218 & 0.86\\
2 & SSS & 7,396 & 29.3\\
3 & STARNET 2.0 & 17,465 & 69.1\\
4 & 2MASS+GSC 1.2 & 182 & 0.72\\
\enddata
\end{deluxetable}

\subsection{Data product and data access}
\label{s:data_product}

Due to  the size  of the catalog,  RAVE DR1  is only accessible  online. The
catalog     fields    are     nevertheless    described     in    Appendix~A, 
Tables~\ref{t:catalog_desc}  to   \ref{t:spec_flag}.  The  catalog   can  be
retrieved      or     queried      from      the     RAVE      collaboration
website.\footnote{http://www.rave-survey.org .}

As the catalog is large, and will  grow in size as further releases are made
public, query  tools are provided on  those websites.  Users  can access the
catalog using  different techniques, either with a  standard query interface
or by field,  using their position on the sky (this  access mode is provided
via  a  clickable  map).   For  DR1  only  tabular  data  are  provided  and
VOtable\footnote{VOtable  is the  standard format  for tabular  data  in the
Virtual Observatory.}   formatted outputs are  offered for ease of  use with
the Virtual Observatory tools.  For the subsequent releases, spectra will be
made  available, requiring  the inclusion  of  more VO  formats (enabling  a
proper description of a spectrum).

In the  field query mode, links  to the Aladin Sky  Atlas \citep{aladin} are
generated,  providing additional  information  in addition  to  VO tools  to
exploit the  data.  However,  this link  to Aladin is  not provided  for the
standard catalog query, as the area on the sky is not restricted.  The first
data  release  catalog,  as  well  as  subsequent  releases,  will  also  be
electronically available at the CDS in the VizieR database~\citep{vizier}.

Again,  DR1  includes  from  the  RAVE survey  only  radial  velocities  and
associated errors, described in  detail above.  Cross-matching with standard
astrometric and photometric catalogs are  provided for ease of use.  Stellar
parameters  are not  part of  this release,  as the  first year  spectra are
contaminated by second order light,  which means that very detailed analysis
is required to extract meaningful values.

\section{Conclusions}

This  first data release  presents radial  velocities for  24,748 individual
stars in the range $9\simlt  I_{DENIS}\simlt 12.5$, obtained from spectra in
the infrared  Calcium-triplet region at  a median resolution of  7,500.  The
total  sky  coverage of  this  catalog  is  $\sim$4,760 square  degrees.  We
demonstrated that  the radial  velocities are not  affected by any  trend in
color  or SNR using  both external  data and  RAVE repeat  observations. The
resulting variance  for each set of  validation data is  consistent with our
estimated errors.

DR1  does  not  include  information  about chemical  abundances  and  other
atmospheric  parameters, for reasons  described above.   The quality  of the
currently acquired spectra  is good enough for derivation  of $T_{\rm eff}$,
gravity,  [M/H], [$\alpha$/Fe],  $V_{\rm rot}\sin  i$,  and micro-turbulence
\citep{fiorucci}, and we expect to  include chemical and atmospheric data in
subsequent data releases.

RAVE  is planned  to observe  until  2010 and  will acquire  up to 1,000,000
spectra.   Incremental  releases,   containing  radial  velocities,  stellar
parameters as  well as spectra, are  now planned on  an approximately yearly
basis, providing an  unprecedented sample of stellar kinematics and chemical abundances in the range
of  magnitudes probing scales between  the  very local  (Hipparcos  based) radial  velocity
surveys (GCS and Famaye et al.~ 2005) and the more distant SDSSII/SEGUE and
surveys with AAOmega, therefore completing our picture of the Milky Way.

\acknowledgments
{\bf Acknowledgments}

We are indebted to the referee Sydney van den Bergh, whose timely and 
justified comments helped us improve the clarity of the manuscript. \\
MS, AS, CB acknowledge financial support from the German Research Foundation (DFG).\\
MS acknowledges financial support from the David and Lucile Packard Foundation.\\
TZ acknowledges financial support from the Slovenian Research Agency.\\
AS is supported by the Alexander von Humboldt Foundation.\\
KCF, QAP, RC, JB-H, BKG, MW are supported from an Australian Research 
Council Grant for the RAVE project.\\
GMS is funded by a Particle Physics and Astronomy Research Council PhD
Studentship.\\
RFGW acknowledges seed money from the  School of Arts and  Sciences at JHU,
plus NSF grant AST-0508996.\\
OB acknowledges financial support from the CNRS/INSU/PNG.\\
EKG is supported by the Swiss National Science Foundation under the
grants 200021-101924 and 200020-106260.\\
AH and MCS acknowledge financial support from NOVA and the Netherlands Organisation for Scientific Research (NWO)\\
JFN and JP's participation in RAVE are supported by Canada's NSERC through a Special Research Opportunities grant\\
HE and MS acknowledge support by the Federal Ministry for Education and Research (BMBF) under FKZ 05AE2EE1/4 and 01AK804A.\\
JF acknowledges support through grants from the W.M. Keck Foundation and the
Gordon and Betty Moore Foundation, to establish a program in data-intensive
science at JHU.\\
The role of the Anglo-Australian
Observatory in providing resources for the first nine months of
observation is gratefully acknowledged.\\
This  research  has made  use  of the  VizieR  catalog  access tool,  CDS,
Strasbourg, France. Data verification is partially based on observations taken at the Observatoire de Haute Provence (OHP) (France), operated by the French CNRS.

\appendix

\centerline{\bf {Appendix A. Catalog description}}

\medskip

This appendix presents the different columns that are provided in the first
data release.

\placetable{t:catalog_desc}
\placetable{t:xid_flag}
\placetable{t:spec_flag}

\begin{deluxetable}{cllcl}
\tablecaption{Description of the RAVE catalog.\label{t:catalog_desc}}
\tablewidth{0pt}
\rotate
\tablehead{
Field Number & Name & Unit & \texttt{NULL} value & Description \\ 
}
\startdata
1 & \textsc{\bf OBJECTID} & - & - & Internal identifier\\
2 & \textsc{\bf RA} & deg & - & Right ascension (J2000)\\
3 & \textsc{\bf DE} & deg & - & Declination (J2000)\\
4 & \textsc{\bf Glon} & deg & - & Galactic longitude\\
5 & \textsc{\bf Glat} & deg & - & Galactic latitude\\
6 & \textsc{\bf RV} & km/s & - & Radial Velocity\\
7 & \textsc{\bf eRV} & km/s & - & Internal radial velocity error\\
8 & \textsc{\bf pmRA} & mas/yr & 9999.9 & Proper motion RA\\
9 & \textsc{\bf epmRA} & mas/yr & 9999.9 & Proper motion error RA\\
10 & \textsc{\bf pmDE} & mas/yr & 9999.9 & Proper motion DE\\
11 & \textsc{\bf epmDE} & mas/yr & 9999.9 & Proper motion error DE\\
12 & \textsc{\bf Spm} & - & - & Source proper motion (see Section~\ref{s:proper_motions})\\
13 & \textsc{\bf Imag} & mag & - & Input catalog $I$ magnitude\\
14 & \textsc{\bf Obsdate} & yyyymmdd & - & Date of observation\\  
15 & \textsc{\bf FieldName}  & - & - & RAVE observations field\\
16 & \textsc{\bf FiberNumber} & -& - & Fiber number on plate\\ 
17 & \textsc{\bf CorrelationCoeff} & - & - & Tonry-Davis correlation coefficient $R$\\
18 & \textsc{\bf PeakHeight} & - & - & Height of the correlation function peak\\
19 & \textsc{\bf PeakWidth} & km/s & - & Width of the correlation function peak width\\
20 & \textsc{\bf CorrectionRV}  & km/s & - & Radial Velocity correction applied\\
21 & \textsc{\bf SkyRV} & km/s & - & Sky radial velocity in fiber \\
22 & \textsc{\bf SkyeRV} & km/s & - & Sky radial velocity error\\
23 & \textsc{\bf SkyCorrelation} & - & - & Tonry-Davis correlation coefficient for sky spectra\\
24 & \textsc{\bf PlateNumber} & - & - & Physical plate number\\
25 & \textsc{\bf SNRatio} & - & - & Signal to Noise ratio\\
26 & \textsc{\bf BT} & mag & 99.99 & $B_{\rm{T}}$ magnitude from Tycho-2\\
27 & \textsc{\bf eBT} & mag & 99.99 & $B_{\rm{T}}$ magnitude error from Tycho-2\\
28 & \textsc{\bf VT} & mag & 99.99 & $V_{\rm{T}}$ magnitude from Tycho-2\\
29 & \textsc{\bf eVT} & mag & 99.99 & $V_{\rm{T}}$ magnitude error from Tycho-2\\
30 & \textsc{\bf USNOID} & - & XXX & USNO-B identifier\\
31 & \textsc{\bf DisUSNO} & mas & 99.999 & Distance to USNO source\\
32 & \textsc{\bf B1} & mag & 99.99 & USNO-B $B1$ magnitude \\
33 & \textsc{\bf R1} & mag & 99.99 & USNO-B $R1$ magnitude \\
34 & \textsc{\bf B2} & mag & 99.99 & USNO-B $B2$ magnitude \\
35 & \textsc{\bf R2} & mag & 99.99 & USNO-B $R2$ magnitude \\
36 & \textsc{\bf IUSNO} & mag & 99.99 & USNO-B $I$ magnitude\\
37 & \textsc{\bf XidQualityFLAGUSNO} & - & X & Cross-identification quality FLAG \mbox{USNO-B} (see table~\ref{t:xid_flag})\\
38 & \textsc{\bf DENISID} & - & XXX & DENIS identifier\\ 
39 & \textsc{\bf DisDENIS} & mas & 99.999 & Distance to DENIS source\\ 
40 & \textsc{\bf IDENIS} & mag & 99.999 & DENIS $I$ magnitude\\ 
41 & \textsc{\bf eIDENIS} & mag & 99.999 & DENIS $I$ magnitude error\\  
42 & \textsc{\bf JDENIS} & mag & 99.999 & DENIS $J$ magnitude\\  
43 & \textsc{\bf eJDENIS} & mag & 99.999 & DENIS $J$ magnitude error\\  
44 & \textsc{\bf KDENIS}  & mag & 99.999 & DENIS $K$ magnitude\\ 
45 & \textsc{\bf eKDENIS}  & mag & 99.999 & DENIS $K$ magnitude error\\ 
46 & \textsc{\bf XidQualityFLAGDENIS} & - & X & Cross-identification quality FLAG DENIS (see Table~\ref{t:xid_flag})\\ 
47 & \textsc{\bf TWOMASSID} & - & XXX & 2MASS identifier\\
48 & \textsc{\bf Dis2MASS} & mas & 99.999 & Distance to 2MASS source\\
49 & \textsc{\bf J2MASS} & mag & 99.999 & 2MASS $J$ magnitude \\
50 & \textsc{\bf eJ2MASS} & mag & 99.999 & 2MASS $J$ magnitude error\\
51 & \textsc{\bf H2MASS}  & mag & 99.999 & 2MASS $H$ magnitude\\
52 & \textsc{\bf eH2MASS}  & mag & 99.999 & 2MASS $H$ magnitude error\\
53 & \textsc{\bf K2MASS}  & mag & 99.999 & 2MASS $K$ magnitude\\
54 & \textsc{\bf eK2MASS}  & mag & 99.999 & 2MASS $K$ magnitude error\\
55 & \textsc{\bf TWOMASSphotFLAG}  & - & XXX & 2MASS photometry flag\\
56 & \textsc{\bf XidQualityFLAG2MASS}  & - & X & Cross-identification quality flag 2-MASS (see Table~\ref{t:xid_flag})\\
57 & \textsc{\bf ZeroPointQualityFLAG} & - & - & zero-point quality flag
(see Table~\ref{t:zeropoint})\\
58 & \textsc{\bf VarFLAG} & - & - & Variability flag, '*' if in GCVS2.0\\
59 & \textsc{\bf SpectraQualityFLAG} & - & - & Spectra quality flag (see Table~\ref{t:spec_flag})\\
\enddata
\end{deluxetable}

\begin{deluxetable}{cc}
\tablecaption{Summary of the cross-identification flag.\label{t:xid_flag}}
\tablewidth{0pt}
\tablehead{
Flag value & Description\\
}
\startdata
A & 1 association within 1~arcsec.\\
B & 2 associations within 1~arcsec.\\
C & More than 2 associations within 1~arcsec.\\
D & Nearest neighbor more than 2~arcsec. away.\\
X & No association found.\\
\enddata
\end{deluxetable}

\begin{deluxetable}{rl}
\tablecaption{Spectra quality flag summary table. The values can be combined to
  give a more accurate description of the spectra.\label{t:spec_flag}}
\tablewidth{0pt}
\tablehead{
\multicolumn{1}{c}{Flag value} & \multicolumn{1}{c}{Description}\\
}
\startdata
a & Asymmetric Ca lines\\
c & Cosmic ray resulting in asymmetric correlation function\\
e & Emission line spectra\\
n & Noise dominated spectra, broad correlation function\\
l & No lines visible, either strong noise or misplaced fiber\\
w & Weak lines, radial velocity can be unreliable\\
g & Strong ghost affecting the wavelength interval used for radial velocity calculation.\\
t & Bad template fit\\
s & Strong residual sky emission\\
cc & Bad continuum\\
r & Red part of the spectra shows problems, noisy\\
b & Blue part of the spectra shows problems, noisy\\
p & Possible binary/double lined\\
\enddata
\end{deluxetable}

\centerline{\bf {Appendix B. External data}}
\medskip

This  appendix presents  the  calibration measurements  obtained  for the  RAVE
survey  during the  first year  of  operation.  Those  calibration data  are
divided in three tables, according  to their source (instrument or catalog).
A     complete    description     of    those     data    is     given    in
Section~\ref{s:external_datasets}.

\begin{deluxetable}{lcccccccccc}
\tablecaption{List of  RAVE targets observed  with the ELODIE  spectrograph. This
    table   presents  the   RAVE  observations   as  well   as   the  ELODIE
    measurements. DENIS  and 2MASS photometry are also  reported. The radial
    velocity error  for RAVE measurements  correspond to the  internal error
    and are not corrected for  zero-point accuracy. Objects flagged with (*)
    are   binary  stars   and   are  discarded   from   the  analysis.   The
    $V_{\rm{T}}$/$I$ column contains the Tycho-2 $V_{\rm{T}}$ magnitudes for
    Tycho-2     objects    and     SSS    $I$     magnitude     for    other
    targets.\label{t:external_ELODIE}}
\rotate
\tablewidth{0pt}
\tablehead{
& \multicolumn{2}{c}{RAVE} & & \multicolumn{2}{c}{ELODIE} & & & & &\\
\cline{2-3}
\cline{5-6}
Identifier & Rad. Vel. & Rad. Vel. & &  Rad. Vel. & Rad. Vel.  & Rad. Vel. & $V_{\rm{T}}$/$I$ & $I_{\rm DENIS}$ & $J\!-\!H$ & $H\!-\!K$\\
& & error & & & error & diff.& & & & \\
& (km/s) & (km/s)& & (km/s) & (km/s) & (km/s) & (mag) & (mag)& (mag) &(mag)\\
}
\startdata
T5027\_00352\_1    &  -7.9& 2.2 & & -8.9 & 1.0 &  1.0& 10.43 & 9.03   & 0.57 & 0.16\\
T5027\_00578\_1    & -32.0& 1.5 & &-32.3 & 1.0 &  0.3& 10.58 & 9.55   & 0.34 & 0.14\\
T5027\_00389\_1    &  39.9& 2.0 & & 39.0 & 1.0 &  0.9& 11.18 & 9.74   & 0.68 & 0.17\\
T5027\_00374\_1    & -47.5& 1.7 & &-43.7 & 1.0 &  0.2& 11.42 & 10.14  & 0.66 & 0.17\\
T5031\_00478\_1(*) & -22.4& 2.3 & &-42.4 & 1.0 & 20.0& 11.26 &\nodata & 0.31 & 0.10\\
C1430054-094720    &  92.7& 2.1 & & 91.9 & 1.0 & -0.2& 11.47 & 11.85  & 0.62 & 0.11\\
C1429286-091608    & -19.2& 1.2 & &-19.2 & 1.0 &  0.0& 10.06 & 10.49  & 0.55 & 0.12\\
\enddata
\end{deluxetable}

\begin{deluxetable}{ccccccccccc}
\tablecaption{2.3m observations of RAVE targets. This table list the RAVE targets
  observed  with the  2.3~m long-slit  spectrograph in  Siding Spring.  RAVE
  measurements, as well as DENIS I magnitude, radial velocity difference and
  2MASS colors are also reported. Stars marked with (*) are variable or
  binary objects and are discarded from the analysis.The estimated radial velocity error for all 2.3~m data is 1.5~km/s.\label{t:external_2p3}}
\rotate
\tablewidth{0pt}
\tablehead{
& \multicolumn{2}{c}{RAVE}& & 2.3~m & & & & & &\\
\cline{2-3}
\cline{5-5}
Identifier & Rad. Vel. & Rad. Vel. & &  Rad. Vel.   &
Rad. Vel. & R value & $V_{\rm{T}}$/$I$ & $I_{\rm DENIS}$ & $J\!-\!H$ & $H\!-\!K$\\
& & error & & & diff.& 2.3~m & \\
& (km/s) & (km/s) & & (km/s) & (km/s) & & (mag) & (mag)& (mag) &(mag)\\
}
\startdata
C1032220-225303    &  103.2 & 2.4 & &  103.6  & -0.4&  43.4 & 11.82  & 12.43 & 0.30 & 0.01\\
C1032220-225303    &  103.8 & 3.4 & &  103.6  & 0.2 &  43.4 & 11.82  & 12.43 & 0.30 & 0.01\\
C1032264-241144    &   52.4 & 2.1 & &   54.3 & -1.9&  75.9 & 11.85  & 12.20 & 0.32 & 0.08\\
C1032264-241144    &   47.9 & 1.5 & &   54.3  & -6.4&  75.9 & 11.85  & 12.20 & 0.32 & 0.08\\
C1033394-215304    &  118.8 & 2.7 & &  120.4  & -1.6&  78.2 & 11.96  & 12.31 & 0.50 & 0.09\\
C1033426-214025(*) &   63.2 & 1.8 & &    5.3  & 57.9&  90.4 & 11.53  & 12.33 & 0.26 & 0.07\\
C1033426-214025(*) &   48.2 & 2.3 & &    5.3  & 42.9&  90.4 & 11.53  & 12.33 & 0.26 & 0.07\\
C1033528-220832    &   34.3 & 2.3 & &   25.2  & 9.1 &  74.8 & 11.99  & 12.33 & 0.22 & 0.10\\
T6073\_00197\_1    &   40.7 & 2.1 & &   38.1  & 2.6 & 124.2 & 11.53  & 9.77  & 0.72 & 0.20\\
T6073\_00197\_1    &   40.0 & 0.6 & &   38.1  & 1.9 & 124.2 & 11.53  & 9.77  & 0.72 & 0.20\\
T6074\_00342\_1    &  -30.3 & 2.1 & &  -22.5  & -7.8& 106.8 & 11.35  & 9.37  & 0.44 & 0.11\\
T6074\_00342\_1    &  -16.6 & 1.1 & &  -22.5  & 5.9 & 106.8 & 11.35  & 9.37  & 0.44 & 0.11\\
T6620\_00749\_1(*) &  -24.4 & 1.9 & &   13.6 &-38.0&  75.2 & 11.77  & 11.10 & 0.26 & 0.09\\
T6620\_00749\_1(*) &  -22.4 & 0.9 & &   13.6 &-36.0&  75.2 & 11.77  & 11.10 & 0.26 & 0.09\\
T6620\_00941\_1    &   86.7 & 2.4 & &   87.8  & -1.1& 132.7 & 11.35  & 10.18 & 0.52 & 0.11\\
T6624\_00025\_1    &   38.3 & 4.3 & &   45.4  & -7.1& 101.7 & 10.71  & 9.80  & 0.52 & 0.17\\
T6624\_00025\_1    &   38.6 & 0.5 & &   45.4  & -6.8& 101.7 & 10.71  & 9.80  & 0.52 & 0.17\\
T6624\_01181\_1    &   22.7 & 1.8 & &   24.2  & -1.5&  90.3 & 10.89  & 10.35 & 0.19 & 0.09\\
T6624\_01181\_1    &   23.2 & 0.7 & &   24.2  & -1.0&  90.3 & 10.89  & 10.35 & 0.19 & 0.09\\
T6637\_00126\_1    &   27.6 & 2.0 & &   31.6  & -4.0& 101.8 & 10.55  & 10.67 & 0.38 & 0.07\\
T6637\_00126\_1    &   26.8 & 0.8 & &   31.7  & -4.8& 101.8 & 10.55  & 10.67 & 0.38 & 0.07\\
C1025107-255418    &  317.1 & 1.7 & &  315.2  & 1.9 & 116.4 & 11.58  & 11.54 & 0.54 & 0.09\\
C1025107-255418    &  317.2 & 1.0 & &  315.2  & 2.0 & 116.4 & 11.58  & 11.54 & 0.54 & 0.09\\
C1025310-260312(*) &   27.0 & 1.7 & &   -1.8  &28.8 & 105.2 & 11.89  & 11.91 & 0.32 & 0.07\\
C1025310-260312(*) &    1.3 & 2.1 & &   -1.8  & 3.1 & 105.2 & 11.89  & 11.91 & 0.32 & 0.07\\
C1026269-255018    &   18.6 & 2.6 & &   15.6  & 3.0 &  65.3 & 11.82  & 11.88 & 0.30 & 0.08\\
C1026269-255018    &   16.5 & 1.8 & &   15.6  & 0.9 &  65.3 & 11.82  & 11.88 & 0.30 & 0.08\\
T6623\_00942\_1    &   16.7 & 1.8 & &   13.2  & 3.5 &  93.4 & 10.06  & 9.42  & 0.36 & 0.15\\
T6623\_00942\_1    &   13.9 & 0.9 & &   13.2  & 0.7 &  93.4 & 10.06  & 9.42  & 0.36 & 0.15\\
T6627\_01266\_1(*) &   -3.0 & 2.6 & &   36.2 &-39.2&  78.9 & 10.87  & 10.26 & 0.17 & 0.11\\
T9317\_01217\_1    &    7.0 & 2.0 & &   11.1  & -4.1&  78.6 & 10.71  & 9.91  & 0.22 & 0.06\\
T9459\_00225\_1    &   -6.0 & 1.1 & &   -7.2  & 1.2 & 103.4 & 10.66  & 10.16 & 0.27 & 0.06\\
T9459\_00391\_1    &    5.5 & 1.8 & &    7.2  & -1.7&  73.3 & 11.22  & 10.10 & 0.63 & 0.09\\
T9459\_00433\_1(*) &  -17.9 & 2.3 & &  -30.7  & 12.8&  53.0 & 10.24  & 9.93  & 0.16 & 0.02\\
T9459\_00433\_1(*) &  -22.0 & 2.1 & &  -30.7  &  8.7&  53.0 & 10.24  & 9.93  & 0.16 & 0.02\\
T9459\_01509\_1    &   45.0 & 1.6 & &   45.1  & -0.1&  72.5 & 11.84  & 10.45 & 0.62 & 0.13\\
T9462\_00133\_1    &   -3.7 & 1.9 & &   -4.5  & 0.8 &  76.5 & 11.34  & 10.30 & 0.43 & 0.15\\
T9462\_01174\_1    &  -17.4 & 1.5 & &  -16.4  & -1.0& 123.1 & 11.86  & 10.14 & 0.61 & 0.24\\
T9463\_00658\_1    &   46.7 & 1.1 & &   43.9  & 2.8 &  91.1 & 10.76  &\nodata& 0.30 & 0.10\\
T9463\_01298\_1    &   35.6 & 1.9 & &   36.3  & -0.7&  75.6 & 11.42  & 10.50 & 0.37 & 0.05\\
T9467\_00768\_1    &   71.8 & 1.6 & &   74.5  & -2.7& 111.4 & 11.57  & 9.79  & 0.71 & 0.21\\
T9318\_00046\_1    &   -0.5 & 1.3 & &    1.1  & -1.6&  81.3 & 11.80  &\nodata& 0.60 & 0.09\\
T9318\_00362\_1    &   15.1 & 1.6 & &   15.8  & -0.7&  92.3 & 10.64  &\nodata& 0.17 & 0.09\\
T9319\_00355\_1    &    2.0 & 1.7 & &    3.5  & -1.5&  96.9 & 10.74  & 9.63  & 0.47 & 0.12\\
T9458\_01749\_1    &   38.2 & 1.0 & &   38.3  & -0.1& 117.1 & 11.52  &\nodata& 0.75 & 0.20\\
T9458\_02394\_1    &  114.4 & 1.6 & &  116.8  & -2.4& 106.3 & 11.61  & 10.23 & 0.73 & 0.11\\
T9459\_00973\_1    &   85.4 & 1.3 & &   86.4  & -1.0&  94.1 & 10.62  & 9.68  & 0.53 & 0.06\\
T9459\_01430\_1    &   -3.7 & 1.4 & &   -1.0  & -2.7&  95.5 & 10.94  & 10.23 & 0.27 & 0.06\\
T9459\_01430\_1    &   -3.2 & 1.8 & &   -1.0 & -2.1&  95.5 & 10.94  & 10.23 & 0.27 & 0.06\\
T9459\_01822\_1    &   24.1 & 1.6 & &   27.0  & -2.9&  91.9 & 11.19  & 9.85  & 0.62 & 0.15\\
T9460\_00265\_1    &   -0.8 & 1.7 & &    0.0  & -0.8&  92.5 & 12.17  & 10.57 & 0.56 & 0.14\\
T9460\_00353\_1    &  -45.4 & 1.4 & &  -42.2  & -3.2&  85.9 & 11.11  & 9.57  & 0.75 & 0.20\\
T9462\_00226\_1    &  -23.3 & 3.0 & &  -25.3  & 2.0 &  68.4 & 11.35  & 10.74 & 0.26 & 0.08\\
T9462\_02202\_1    &   27.3 & 1.5 & &   23.7  & 3.6 & 102.2 & 10.35  & 9.85  & 0.29 & 0.06\\
T9463\_01633\_1    &   16.0 & 3.8 & &   16.4 & -0.4&  74.2 & 12.33  & 11.47 & 0.39 & 0.04\\
T9464\_00658\_1    &    1.3 & 1.0 & &   -0.7  & 2.0 &  80.7 & 10.82  & 10.09 & 0.22 & 0.07\\
T9464\_00658\_1    &    0.4 & 1.5 & &   -0.7  & 1.1 &  80.7 & 10.82  & 10.09 & 0.22 & 0.07\\
T9464\_00865\_1    &    8.3 & 1.5 & &    3.2  & 5.1 &  67.2 & 10.76  & 9.96  & 0.33 & 0.15\\
T9467\_00478\_1    &   23.1 & 6.6 & &   22.4  & 0.7 &  98.9 & 11.80  & 10.24 & 0.67 & 0.22\\
T9317\_01143\_1    &   13.5 & 2.8 & &   19.7  & -6.2& 100.6 & 11.69  & 10.48 & 0.55 & 0.14\\
T9317\_01143\_1    &   17.5 & 1.6 & &   19.7 & -2.2& 100.6 & 11.69  & 10.48 & 0.55 & 0.14\\
T9458\_00843\_1    &    9.7 & 1.2 & &    8.6  & 1.1 &  87.8 & 10.27  & 9.50  & 0.35 & 0.11\\
T9458\_00934\_1    &  -41.9 & 1.3 & &  -41.0  & -0.9& 112.2 & 12.10  & 10.97 & 0.59 & 0.11\\
T9460\_01761\_1    &   31.1 & 1.1 & &   34.6  & -3.5&  80.0 & 11.03  & 10.34 & 0.22 & 0.09\\
T9462\_01690\_1    &    0.3 & 2.7 & &    2.2  & -1.9& 111.6 & 11.35  & 9.83  & 0.58 & 0.16\\
T9317\_00415\_1    &   22.3 & 2.4 & &   20.3  & 2.0 & 116.1 & 11.43  & 10.34 & 0.48 & 0.13\\
T9458\_02020\_1    &   23.2 & 1.9 & &   17.0  & 6.2 &  58.5 & 9.49   & 9.28  &-0.04 & 0.12\\
T9460\_00219\_1    &   75.3 & 1.2 & &   75.2  & 0.1 & 157.2 & 11.71  & 10.17 & 0.59 & 0.16\\
T9462\_01219\_1(*) &   57.5 &10.4 & &   30.4  &27.1 &  88.9 & 9.91   &  9.54 & 0.10 & 0.08\\
T9464\_00050\_1    &   28.9 & 0.9 & &   25.6  & 3.3 & 100.4 & 10.69  & 10.15 & 0.20 & 0.06\\
C1519196-191359    & -407.3 & 2.3 & & -409.0  & 1.7 &  76.8 & 11.32  & 12.12 & 0.66 & 0.14\\
C1508217-085010    & -289.8 & 3.0 & & -287.8  & -2.0&  72.7 & 11.29  & 11.91 & 0.49 & 0.07\\
C1536201-144228    & -329.2 & 2.8 & & -330.4  & 1.2 &  46.4 & 11.82  & 12.42 & 0.49 & 0.16\\
T7274\_00734\_1    &  407.9 & 1.5 & &  407.8  & 0.1 & 108.2 & 10.06  &\nodata& 0.46 & 0.07\\
C1905210-751503    & -153.1 & 3.6 & & -148.9  & -4.2&  79.6 & 11.69  & 12.00 & 0.64 & 0.14\\
T8395\_01513\_1    & -296.3 & 2.0 & & -295.2 & -1.1& 140.4 & 11.30  & 9.99  & 0.51 & 0.05\\
T6671\_00389\_1    &    0.7 & 2.0 & &   -5.0  & 5.7 &  79.0 & 10.80  & 10.14 & 0.26 & 0.10\\
C1435432-164545    &  -15.1 & 3.7 & &  -17.2  & 2.1 &  40.4 & 11.51  & 12.16 & 0.57 & 0.10\\
C1254296-164722    &   71.7 & 2.2 & &   74.3  & -2.6&  44.0 & 10.82  &\nodata& 0.38 & 0.08\\
C1532041-135910    &   44.2 & 1.4 & &   36.8  & 7.4 &  43.5 & 11.81  & 12.09 & 0.45 & 0.09\\
C2030129-660620    &   91.6 & 2.3 & &   91.9  & -0.3&  63.6 & 11.54  & 12.09 & 0.37 & 0.08\\
T7270\_00796\_1    &  151.4 & 1.5 & &  152.4  & -1.0&  81.1 & 11.88  &\nodata& 0.44 & 0.08\\
T7524\_00065\_1    &  353.2 & 1.2 & &  353.8  & -0.6& 130.1 & 11.37  &\nodata& 0.55 & 0.14\\
T7535\_00160\_1    &  356.6 & 1.4 & &  358.8  & -2.2&  99.7 & 11.43  & 10.22 & 0.56 & 0.08\\
T9527\_00088\_1    & -102.2 & 2.1 & &  -89.6  &-12.6&  89.9 & 11.99  & 10.57 & 0.62 & 0.13\\
C0314269-375257    &   69.3 & 1.8 & &   72.9  & -3.6&  83.8 & 11.97  & 12.38 & 0.48 & 0.13\\
C2118490-174605    & -125.9 & 3.1 & & -133.2  &  7.3&  67.7 & 11.54  & 12.37 & 0.66 & 0.15\\
C2234046-564051    &  -12.0 & 2.0 & &  -15.2  &  3.2&  77.8 & 11.88  & 12.05 & 0.51 & 0.12\\
C2330284-410842    &   51.6 & 2.0 & &   48.8  &  2.8&  92.6 & 11.36  & 11.51 & 0.46 & 0.07\\
T7006\_01317\_1    &   82.6 & 2.8 & &   81.5  &  1.1&  77.3 & 11.68  &\nodata& 0.39 & 0.03\\
\enddata
\end{deluxetable}

\begin{deluxetable}{lcccccccccc}
\tablecaption{List of  RAVE targets in  the Geneva-Copenhagen survey.  This table
presents    the    Geneva-Copenhagen     data    together    with    RAVE
measurements. DENIS and 2MASS photometry are reported for
convenience. Objects flagged with (*) are binary stars and are discarded from
the analysis.\label{t:external_Copenhagen}}
\rotate
\tablewidth{0pt}
\tablehead{
& \multicolumn{2}{c}{RAVE} & & \multicolumn{2}{c}{GENEVA} & & & & &\\
\cline{2-3}
\cline{5-6}
Identifier & Rad. Vel. & Rad. Vel.  & &  Rad. Vel. & Rad. Vel.  &
Rad. Vel. & $V_{\rm{T}}$/$I$ & $I_{\rm DENIS}$ & $J\!-\!H$ & $H\!-\!K$\\
& & error & & & error & diff.& & & & \\
& (km/s) & (km/s) & & (km/s) & (km/s) & (km/s) & (mag) & (mag)& (mag) &(mag)\\
}
\startdata
T8468\_01019\_1    & 8.7   & 1.3 & & 8.2   & 0.3 & 0.5 & 9.95   &        & 0.36 & 0.10\\
T6053\_00177\_1    & 57.9  & 1.4 & & 55.4  & 0.7 & 2.5 & 11.56  & 10.39  & 0.59 & 0.10\\
HD 143885          & 1.7   & 1.3 & & -0.3  & 0.2 & 2.0 & 7.66   & 8.82   & 0.24 & 0.07\\
HD 146124          & 14.2  & 2.2 & & 10.8  & 0.2 & 3.4 & 6.85   &        & 0.31 & 0.14\\
HD 153713          & -4.7  & 1.8 & & -6.8  & 0.3 & 2.1 & 8.54   & 9.04   & 0.45 & 0.10\\
HD 154550(*)       & -7.3  & 1.7 & & -7.4  &     & 0.1 & 8.18   & 8.77   & 0.24 & 0.03\\
HD 155221          & -49.3 & 1.4 & & -49.7 & 0.2 & 0.4 & 7.80   & 9.17   & 0.31 & 0.11\\
HD 155755          & -0.3  & 1.5 & & -2.4  & 0.4 & 2.1 & 7.91   &        & 0.23 & 0.07\\
HD 156741(*)       & -51.4 & 1.3 & & -52.5 & 0.2 & 1.1 & 8.39   & 8.67   & 0.35 & 0.09\\
HD 157316          & -1.9  & 1.9 & & -1.3  & 0.9 & -0.6& 6.26   & 8.80   & 0.19 & 0.05\\
HD 157387          & 24.5  & 1.1 & & 24.8  & 0.3 & -0.3& 7.25   & 7.61   & 0.21 & 0.06\\
HD 157887          & -20.7 & 1.2 & & -20.7 & 0.2 & 0.0 & 7.67   & 8.44   & 0.29 & 0.08\\
T8454\_00006\_1    & 7.8   & 1.5 & & 4.3   & 0.2 & 3.5 & 10.18  &        & 0.38 & 0.08\\
\enddata
\end{deluxetable}
\end{document}